# High bandwidth temporal RF photonic signal processing with Kerr micro-combs: integration, fractional differentiation and Hilbert transforms

Mengxi Tan, Xingyuan Xu, * Jiayang Wu, and David J. Moss

Optical Sciences Centre, Swinburne University of Technology, Hawthorn, VIC 3122, Australia. (dmoss@swin.edu.au).
* X. Xu current address: Electro-Photonics Laboratory, Dept. of Electrical and Computer Systems Engineering, Monash University, VIC3800, Australia.

*Abstract*— **Integrated Kerr micro-combs, a powerful source of many wavelengths for photonic RF and microwave signal processing, are particularly useful for transversal filter systems. They have many advantages including a compact footprint, high versatility, large numbers of wavelengths, and wide bandwidths. We review recent progress on photonic RF and microwave high bandwidth temporal signal processing based on Kerr micro-combs with spacings from 49-200GHz. We cover integral and fractional Hilbert transforms, differentiators as well as integrators. The potential of optical micro-combs for RF photonic applications in functionality and ability to realize integrated solutions is also discussed.**

## I. INTRODUCTION

All-optical signal processing based on nonlinear optics has proven to be extremely powerful, particularly when implemented in photonic integrated circuits based on highly nonlinear materials such as silicon [1-3]. All optical signal processing functions include all-optical logic [4], demultiplexing at ultra-high bit rates from 160Gb/s [5] to over 1Tb/s [6], optical performance monitoring (OPM) using slow light [7,8], all-optical regeneration [9,10], and others [11-16]. Complementary metal oxide semiconductor (CMOS) compatible platforms are centrosymmetric and so the $2^{rd}$ order nonlinear response is zero. Hence, nonlinear devices in these platforms have been based on the $3^{rd}$ order nonlinear susceptibility including third harmonic generation [11,17-21] and the Kerr nonlinearity ($n_2$) [1,2]. The efficiency of Kerr nonlinearity based all-optical devices depends on the waveguide nonlinear parameter ($\gamma$). Although integrated silicon nanowires can achieve extremely high nonlinear parameters ($\gamma$), they suffer from high nonlinear optical loss due to two-photon absorption (TPA) and the ensuing free carriers [2]. Even with the use of p-i-n junctions to sweep out the carriers , the nonlinear figure of merit (FOM = $n_2$ / ($\beta\,\lambda$), where $\beta$ is the TPA and $\lambda$ the wavelength) of silicon is only 0.3 in the telecom band - too low to achieve good performance. While TPA can be used as a positive tool for some nolinear functions [22-24], for most processes the low FOM in the telecom band poses a problem. This has motivated research on other nonlinear optical platforms including chalcogenide glass [25-34]. However, although these platforms may offer advantages, most are not compatible with CMOS fabrication which is a significant consideration for practical and cost effective manufacturing.

Around 2007, new platforms for nonlinear optics were reported that are CMOS compatible with negligible TPA in the telecom band, including  silicon nitride [35, 36] and high-index doped silica glass (Hydex) [37-47]. In addition to no nonlinear absorption, these platforms displayed a reasonable Kerr nonlinearity that yielded an extremely high nonlinear FOM as well as a nonlinear parameter high enough to sustain substantial parametric gain.  Following the first report of a micro-ring resonator (MRR) based frequency comb source, driven by the Kerr optical nonlinearity in 2007 [48], the first fully integrated optical parametric oscillators were reported in 2010 [36, 37] in these new CMOS compatible platforms. Since 2010 the field of integrated micro-combs, or "Kerr combs" has become one of the largest fields in optics and photonics [47]. Integrated optical Kerr frequency comb sources, or "micro-combs" are a new and powerful tool to accomplish many new functions on an integrated chip, due to their very high coherence, while also offering flexible control wavelength spacing. Optical micro-combs are produced via optical parametric oscillation driven by modulational instability, or parametric, gain in integrated ring resonators, offering huge advantages over more conventional multiple wavelength sources. Many breakthroughs have been reported with Kerr micro-combs, from innovative mode-locked lasers [49-52] to quantum optical photonic chips [53-61], ultrahigh bandwidth optical fiber data transmission [62-64], optical neural networks [65-67], integrated optical frequency synthesizers [68]. Micro-combs have been extensively reviewed [47, 69 - 76]. The success of these new CMOS platforms motivated the search for even higher performing CMOS compatible platforms including as amorphous silicon [77] and silicon rich silicon nitride [78], searching for the combination of low linear and nonlinear loss together with a high nonlinearity.

All-optical signal processing based on nonlinear optics has attracted significant interest over the years for its ability to achieve ultrahigh bandwidth operation without optical to electronic (or visa versa) conversion, for signal processing functions for telecommunications and RF/microwave systems. RF photonic applications include radar systems to signal generation and processing [79-128], and are attractive because of the ultra-high bandwidths as well as low transmission loss and strong immunity to electromagnetic interference. There are different RF photonic approaches including methods that map the optical filter response onto the RF domain. This is perhaps best represented by integrated devices based on stimulated Brillouin scattering [88-95], that have achieved high RF resolution — down to 32 MHz, a stopband discrimination > 55 dB. A key approach to reconfigurable transfer functions for signal processing has been transversal filter methods [96-100]. These operate by generating progressively delayed and weighted replicas of an RF signal multicast onto many optical carriers, which are subsequently summed via photo-detection. Transverse filters can achieve a wide range of RF functions solely by varying the tap weights, and so this approach is very attractive for advanced dynamically adaptive RF filters. Discrete diode laser arrays [101] and fibre and integrated Bragg grating arrays and sampled gratings [103] have been used to generate the required taps. However, while offering advantages, these methods have increased complexity and footprint, limiting performance due to the limited number of wavelengths. Alternative methods such as electro-optic (EO) or acousto-optic (AO) combs [102,104,105], can help overcome this, but they require many high bandwidth modulators and high-frequency RF sources.

Kerr micro-combs were invented about 10 years ago [36, 37] and have been very successful. They provide advantages over other sources of multiple wavelengths for RF devices. They have achieved extremely high bandwidth data communications as well as a wide range of microwave signal processing devices [107-128]. Their combs spacings can be much wider than electro-optic combs, and in many ways EO and micro combs are complementary. EO combs excel at finer spacings from 10's of megahertz to 10 - 20 GHz, while integrated micro-combs typically have much wider spacings from 10's of GHz to 100's of GHz and even THz. Larger comb spacings have much wider Nyquist zones for large RF bandwidth operation, whereas smaller spacings provide many more wavelengths or RF "taps", although at the expense of a smaller Nyquist zones. Micro-combs provide more wavelengths while still with a large FSR, all in a small footprint. For RF transversal filters the number of wavelengths determines the number of channels for RF true time delays and RF filters [85, 121]. Systems such as beamforming devices [112] can also be improved in quality factor and angular resolution. Other approaches to filtering include RF bandwidth scaling [125] that yields a particular bandwidth for each wavelength channel, with the total bandwidth (maximum RF signal bandwidth) will depend on the number of wavelengths which is dramatically increased with micro-combs.

Recently [121], we reviewed transversal filtering and bandwidth scaling methods based on Kerr micro-combs, as applied to RF and microwave spectral filters. In this paper, we review microwave and RF high bandwidth temporal signal processing based on Kerr micro-combs. We cover both integral and fractional order Hilbert transformers and differentiators, as well as RF integrators. We discuss the trade-offs between wide spaced micro-combs with an FSR of 200GHz [109-111] with record low FSR micro-combs with a spacing of 49 GHz, operating via soliton crystals [122-128]. We highlight their potential and future possibilities, contrasting the different methods and use of the differently spaced micro-combs. While 200GHz Kerr micro-combs have been successful for RF transversal filters, allowing high versatility and dynamic reconfigurability, their large comb spacing limits the number of wavelengths to typically < 20 in the C-band. This is an important consideration because transversal filters need many components such as optical amplifiers and spectral shapers that are only available at telecom wavelengths (1530-1620nm). This limitation in the number of wavelengths has limited the frequency resolution, bandwidth, and dynamic reconfigurability of micro-comb based RF filters. To overcome this, we focused on RF transversal filters with record high numbers of wavelengths − up to 80 in the C-band [117]. This represents the highest for micro-comb based RF filters, and is a result of a record low spacing of 49GHz for the Kerr micro-comb. This resulted in RF filters with [121] $Q_{RF}$ factors bandpass filters what were 4 times higher than the 200GHz combs yielded. In turn, for time dependent signal processing this results in a significant improvement in both reconfigurability and bandwidth. Our results confirm the feasibility of achieving high performance reconfigurable transversal RF filters for signal processing with reduced footprint, complexity, and cost.

## II. INTEGRATED KERR MICRO-COMBS

The formation of micro-combs is complex and arises from a combination of a high nonlinear parameter together with low nonlinear and linear loss, and finally with careful dispersion engineering. Many different material platforms have been reported for micro-combs [47] including silica, magnesium fluoride, silicon nitride, and doped silica glass [47, 70, 85]. In 2008 [39] we reported efficient FWM at low (milliwatt) powers with a 575GHz FSR spaced comb, generated by a ring resonator with a relatively low Q-factor of about 60,000. This was the first report of low power continuous wave (CW) nonlinear optics in a silica glass based platform, and in 2010 was followed by the first report of an integrated micro-comb [37, 38]. These were realized in Hydex and silicon nitride and were inspired by the first micro-combs in toroid resonators [48]. A significant development came [116, 117] with integrated micro-combs having record low FSR's of < 50GHz, significantly increasing the number of available wavelengths to over 80 in the telecom band. Apart from their low spacing, these micro-combs operated via a different process than DKS states [66-73], a process termed soliton crystals [129, 130]. Many breakthroughs have been achieved with micro-combs, including ultralow pump power combs [131], dark solitons [132], laser-cavity solitons [133] and others [134-139].

The microcombs that formed the basis of the research discussed here were fabricated in Hydex glass [37, 38], a CMOS compatible platform. Ring resonators with Q's from 60,000 to greater than 1.5 million have been reported, with radii from 592 - 48 μm, corresponding to FSRs from 49 - 575 GHz. The RF signal processors reviewed here were based on combs with spacings of 200GHz (Fig. 1a) and 49GHz (Fig. 1b). Hydex glass layers ($n = \sim1.7$ at 1550 nm) were deposited via PECVD (plasma enhanced chemical vapour deposition), and patterned via deep UV stepper mask aligner photolithography. The waveguides were reactive ion etched to create waveguides with very low surface roughness. An upper cladding layer of silica glass ($n = \sim1.44$ at 1550 nm) was then deposited. We typically use a vertical coupling scheme between the bus and ring resonator, with a gap of about 200nm that can be controlled by film growth - more accurate than lithography. The advantages Hydex include its very low linear loss ($\sim0.06$ dB·cm$^{-1}$), its reasonable nonlinear parameter (0.233 W$^{-1}$·m$^{-1}$), and negligible TPA even up to many GW·cm$^{-2}$. We reported high Q's $\sim$1.5 million (Fig. 1) for both the 49GHz and 200GHz FSR MRRs. After packaging with fiber pigtails, the coupling loss was about 0.5 - 1 dB per facet, achieved with on-chip mode converters.

To generate combs with the 200GHz devices, the CW pump power was amplified to > +30 dBm and the wavelength was tuned from blue to red near a TE resonances at ~1550 nm. When the difference between the pump and cold resonance wavelength became sufficiently small it resulted in the intracavity power reaching threshold. At this stage, modulation instability gain resulted in oscillation [47], generating primary combs with a spacing given by the MI gain peak — a function of the dispersion and intra-cavity power. As the detuning was varied, single FSR spaced micro-combs appeared. While these states were not solitons, we found that operating in single soliton states such as the dissipative Kerr solitons (DKS) [66] was in fact not necessary. This is significant since, while much is now understood about DKS solitons and significant progress has been made [139], DKS states require complicated pump tuning dynamics with simultaneous amplitude and wavelength sweeping, including in reverse directions, to be able to "kick" the solitons out of chaotic states. Our early work on micro-combs was based on the 200GHz FSR combs that operated in this partial coherence. While not rigorous soliton states, they were still low noise and avoided the chaotic regime [47] - we found them to be sufficient for our RF work. While their spectra (Figs. 1b,c) indicate that they are not solitons, nonetheless they successfully demonstrated many functions for RF signal processing [85, 107, 109 -111].

More recently, we have used microcombs featuring a record low spacing of 49GHz, based on soliton crystals [129, 130]. These arise from mode crossings and are easier and more robust to generate than other solitons, and even partially coherent states. They can be generated without complex pump dynamics, and indeed can be achieved even via manual control of the pump wavelength. The reason for this lies in the fact that the internal optical energy in the cavity of the soliton crystals is very close to the chaotic state, and so when they are generated from chaos there is only a small change in optical cavity energy, resulting in almost no shift in resonant wavelength. It is the shift in resonant wavelength and internal energy that makes DKS states challenging to generate. This also greatly increases the efficiency of the soliton crystals (comb line relative to the pump energy) relative to DKS states (single solitons in particular). However, the compromise with soliton crystals is that their spectra are not flat – they have nonuniform "curtain" patterns. While this can require spectral flattening, it has not prevented them from achieving many high performance microwave and RF functions. Soliton crystals are based on a mode-crossing in the resonator and so this needs to be engineered, as well as anomalous dispersion, but this has not posed a significant difficulty and high fabrication yield is achievable [64].

When generating soliton crystals, we typically tuned the pump laser wavelength manually across a resonance (TE polarized). When the pump (with sufficient power) was aligned well enough with the resonance, a primary comb was generated, similar to the 200GHz states. In this case, however, as the detuning was changed further, distinctive 'fingerprint' optical spectra were observed (Fig. 2), arising from spectral interference between tightly packed solitons in the cavity – solitons that formed the soliton crystals [129, 130]. We observed a small abrupt jump in the intracavity power (Fig. 2(b)) when these spectra appeared. At the same time we observed a large reduction in the RF intensity noise (Fig. 2(c)). These observations indicate soliton crystal formation [38], although to prove this one would need to perform time-resolved autocorrelations. The specific micro-comb states reached was not important for RF applications, only that low RF noise and high coherence be achieved, and we found this to be easy to generate with simple pump wavelength tuning. Soliton crystals yielded the lowest noise of all micro-comb states, and we have focused on these for our microcomb work, including a low phase-noise microwave oscillator [120].

## III. RF TRANSVERSAL FILTER: THEORY AND EXPERIMENT

The transfer function of a general transversal structure can be described as [85]

$$F(\omega) = \sum_{k=0}^{M-1} h_k e^{-j\omega KT} \tag{1}$$

Where $j = \sqrt{-1}$, $\omega$ is the angular frequency of the RF signal, $M$ is the tap number, $T$ is the time delay between taps, $h_k$ is the $k_{th}$ tap coefficient for the discrete impulse response of $F(\omega)$, obtained performing the inverse Fourier transform of $F(\omega)$, temporally windowed with a cosine bell. The transversal filter is equivalent to a finite impulse response digital filter. Both are fundamental tools for RF signal processing. By designing the tap weights $h_k$ for each wavelength, different signal processing transfer functions can be obtained including integrators, differentiators, Hilbert transforms and more. The response function Nyquist frequency (maximum RF frequency) is given by $f_{Nyquist} = 1/2T$. Figure 3 shows the transversal filter and multiple wavelength source. First, the RF signal is multicast onto all microcomb wavelengths with a modulator, followed by transmission through a dispersive device to generate a wavelength-dependent delay. Second-order dispersion, yielding a linear relationship between wavelength and delay, is used to progressively delay the replicas. These delayed signals can be either separately converted into electronic RF signals with wavelength demultiplexers and photodetector arrays for applications such as RF true time delays [116], or summed by photodetection for RF transversal filters signal processing [85]. The large number of wavelengths from the micro-comb has major advantages since the number and amplitude of progressively delayed RF replicas, or taps, determines the performance of the system. By setting the tap coefficients, a reconfigurable transversal filter with virtually any transfer function can be realized, including both integral and fractional differentiators, [109,127] Hilbert transformers [107, 126], integrators [123], waveform generators, [128] bandpass filters [111,117, 121], and more.

Figure 4 shows the transversal filters for the 200GHz (Fig 4a) and 49GHz (Fig. 4(b)) combs. In both cases the combs are amplified and then processed by one (200GHz case) or two (49GHz case) waveshapers to flatten then shape the spectra. A second waveshaper was used for the 49GHz FSR comb to pre-flatten the spectra of the soliton crystals. The power of each comb line was controlled by the waveshaper according to the tap weights. To increase the accuracy, real-time feedback control was used to shape the comb line power. The comb lines were then spatially divided into two according to the sign of the tap coefficients – one path for negative weights and one for positive. Next, for the 200GHz micro-comb experiments the signal was passed through a 2×2 balanced Mach-Zehnder modulator (MZM) biased at quadrature which simultaneously modulated the input RF signal on both positive and negative slopes, yielding replicas of the signal with phase and tap coefficients having both signs. For 49GHz FSR experiments, both positive and negative taps were achieved by spatially separating the wavelengths according to the sign of the tap coefficients and then feeding them into the positive and negative input ports of a balanced photodetector (Finisar BPDV2150R) (Figure 4b).

For the 200GHz experiments, the signal was modulated by the MZM and went through ~2.122-km of standard single mode fibre (SMF, ~17.4 ps/nm/km), yielding a delay $T$ ~59 ps between wavelengths, or taps (the channel spacing of the time delay lines equalled the micro-comb FSR), resulting in a Nyquist frequency of ~8.45 GHz. The bandwidth of the systems (Nyquist frequency) could be increased by decreasing the time delay which, because of the large FSR of the compact MRR, could reach beyond 100 GHz. Finally, the weighted and delayed taps were combined by a high-speed photodetector (Finisar, 40 GHz bandwidth) into electronic RF signals.

For the 49GHz FSR microcomb based transversal filter, the signal went through 5km of SMF to generate the delay taps. The fibre dispersion was identical (17 ps/nm/km) in the 200GHz system, but the time delay between adjacent taps was different at $T \sim 34.8$ ps. This resulted in a bandwidth (Nyquist frequency, half of $FSR_{RF}$) of 14.36 GHz. This can be increased by decreasing the time delay (with shorter fiber), although with reducing the tuning resolution. The transversal filter bandwidth was limited by the comb spacing. For the 49 GHz comb, significant crosstalk between adjacent wavelength channels occurred for RF frequencies > 24.5 GHz. This could be remedied by using a micro-comb with a larger spacing, although at the expense of yielding fewer comb lines over the C-band. Although standard optical fibre is often used to generate the delays, this can be done more efficiently with other methods. Typically, only 2-km of SMF, with a dispersion of 34 ps/nm, is needed. This is within the range of multi-channel tunable dispersion compensators [140-143], which can also be designed with a built-in dispersion slope offset. This would not only yield a compact and latency free delay but enable tunability to adjust the Nyquist zone. In the following sections, we review progress made in high bandwidth temporal RF and microwave signal processing functions beginning with Hilbert transforms, followed by differentiation and then integration, all based on Kerr micro-combs.

## IV. HILBERT TRANSFORMS

The Hilbert transform (HT) is a fundamental signal processing operation with wide applications to single sideband modulators, radar systems, signal sampling, speech processing, measurement systems, and more [144]. Integral HTs perform a $\pm 90°$ phase shift around a central frequency, with an all-pass amplitude transmission. Fractional Hilbert transforms (FHTs) that yield a variable phase shift represent a powerful extension to the standard HT, with applications to secure single sideband communications [145], hardware keys [145, 146], and in forming images that are edge enhanced relative to the input object, where one can select the edges that are enhanced as well as the degree of edge enhancement [147]. Electronic fractional Hilbert transformers are limited in bandwidth [145, 148], while photonic approaches yield wide bandwidths and a high level of EMI immunity. HTs using free-space optics [146, 149] have been reported that have yielded high levels of performance but these devices are complicated and large in size. Phase-shifted fibre Bragg grating based Hilbert transformers typically have bandwidths of a few 100GHz [150-153] but only yield a precise FHT for signals with fixed fractional orders, specific bandwidths, and do not operate on the RF signal itself but the complex optical field. This also holds for integrated reconfigurable microwave processors [154] as well Bragg grating approaches [155]. In practice, it is not the operation on the complex optical field that is of interest for RF signal reshaping and measurement but the operation on the actual RF and microwave signals [156-163].

Integrated Hilbert transformers realized using micro-ring or micro-disc resonators, Bragg gratings in silicon, and integrated InP-InGaAsP photonic chips yield compact devices with high stability and mass-producibility [154, 155, 159-161]. Nevertheless, they only operate on the complex optical field and generally only provide a limited phase shift tuning range. Transversal filter approaches offer high reconfigurability [162, 163] but need many discrete lasers, thus increasing system cost, complexity, and size, and restricting the tap number and hence performance. Optical frequency combs provide a single high-quality source that can generate many wavelengths. These include mode-locked lasers of various types [164], modulators [165], as well as resonators [166]. Of these, Kerr micro-combs generated by micro-resonators provide many wavelengths with greatly reduced footprint.

A fractional Hilbert transformer has a transfer function in the spectral domain given by [80, 126]:

$$H_P(\omega) = \begin{cases} e^{-j\varphi}, & if\ 0 \leq \omega < \pi \\ e^{j\varphi}, & if\ -\pi \leq \omega < 0 \end{cases} \tag{2}$$

where $j = \sqrt{-1}$, $\varphi = P \times \pi / 2$ is the phase shift, and $P$ denotes the fractional order. From Eq. (2), a FHT can be viewed as essentially a phase shifter with a phase shift of $\pm \varphi$ centered about a frequency $\omega_c$. This defaults to a standard integral HT when $P = 1$. The impulse response is a continuous hyperbolic function:

$$h_P(t) = \begin{cases} \frac{1}{\pi t}, t \neq 0 \\ \cot(\varphi), t = 0 \end{cases} \tag{3}$$

For digital transforms, this hyperbolic function gets truncated and temporally sampled with the discrete taps. The sample time step $\Delta t$ yields the null frequency $f_c = 1/\Delta t$. The order of the FHT can be tuned continuously by only varying the tap coefficient at $t = 0$ without varying any of the other tap coefficients [146].

To achieve a Hilbert Transform the normalized power of each comb line is given by:

$$p_n = \frac{1}{\pi |n - N/2 + 0.5|} \qquad (4)$$

where $N$ is the number of or filter taps or comb lines, and $n = 0, 1, 2, \ldots, N\text{-}1$ is the comb index.

In 2015, the first Kerr micro-comb based Hilbert transformers was demonstrated, that employed as many as 20 wavelengths [107] from a microcomb that had an FSR of 200GHz. This device achieved a record-high RF range of 5 octaves. Figures 5(a) and 5(b) show the shaped optical micro-combs measured with an optical spectrum analyzer for combs having 20 and 12 wavelengths, respectively. The designed wavelength target powers, or tap weights, are shown (green crosses) in Figs. 5(a) and 5(b). The waveshaper shaped the comb line powers to within +/-0.5 dB of the designed weights. This was feasible  since the waveshaper had a resolution of 10 GHz much less than the comb-line spacing of 200GHz. After the comb line powers were adjusted to give the correct weights needed for the HT impulse response, with the un-needed comb lines extinguished below the noise floor, the RF frequency amplitude and phase response was measured using a vector network analyzer (VNA).

Figure 6a shows the photonic HT filter RF amplitude response in frequency for 12, 16 and 20 taps. All measurements agree with theory, showing very low (< 3 dB) amplitude ripple. All 3 filters were measured to have the same null frequency at 16.9 GHz, corresponding a tap spacing of $\Delta t = 1/f_c = 59$ ps, matching the delay difference between comb lines. This difference was equal to the *FSR* of 1.6 nm, and was generated after being transmitted through spool of 2.122-km SMF with a dispersion of $D = 17.4$ ps/nm/km. The null frequency could be adjusted by employing different lengths of fiber to adjust the delay tap spacing, or alternatively by using tunable dispersion devices [140-143]. Increasing the tap number increases the filter bandwidth so that with 20-taps, the Hilbert transformer had a 3-dB bandwidth from 0.3 - 16.4 GHz - more than five octaves - which was a record for any HTs. The bandwidth can be increased even further using more comb lines. Indeed, only a small fraction of the generated comb lines was actually employed for these filters. The number of taps was actually limited by the waveshaper, which was a C-band device, as well as the EDFA bandwidth. Amplitude ripple over the passband can be reduced by apodizing the tap coefficients. Figure 6(b) shows the measured phase response as a function of the number of taps,  showing a very similar response. Each shows a near constant phase close to -90° over the passband. There are some small variations compared with the ideal -90° phase near zero frequency, and particularly for the exact null frequency $f_c = 16.9$ GHz.

Recently, [126] Tan *et. al* used a Kerr microcomb having a much lower comb FSR of 49GHz that provided many more lines (80) across the C band, as the basis of a fractional Hilbert transformer (FHT). FHTs were achieved with phase shifts of 15°, 30°, 45°, 60°, 75°, and 90°, equivalent to fractional orders (0.17, 0.33, 0.5, 0.67, 0.83, 1.0) by shaping the comb lines with the required weights,. Up to 17 wavelengths were used - the space between the 8-9th and 9-10th lines was 0.8 nm, with the remaining being 1.6 nm, chosen to maximize the Nyquist zone. A 2.1-km length of single mode fibre ($\beta = \sim17.4$ ps / nm / km) provided a delay line between channels of $\tau = L \times \beta \times \Delta\lambda = \sim29.4$ ps, consistent with a FSR$_{RF}$ of $1/2\tau = \sim17$ GHz. This bandwidth can be varied (increased) by changing the length (shortening) of fibre.

The theoretical amplitude and phase response of the FHTs for all phase shifts are given in Figs. 7 (a)-(f) versus tap number. Fig. 8(a) and (b) show the RF frequency phase and amplitude response of a FHT with phase shift 45° and 5, 9, 13 and 17 taps. Figure 8c) shows the bandwidth (3dB) vs tap number where it is seen that the bandwidth increases with tap number. With 17 taps, the base-band RF bandwidth was 480 MHz to 16.45 GHz - more than 5 octaves – with a very small variation of ±0.07 rad in the phase within the band - a record performance for FHTs. For a standard 90° HT the micro-comb generated > 40 taps, reducing the amplitude ripple and root-mean-squared error (RMSE) across the passband [126]. Fig. 8 (c) shows that the theoretical bandwidth rapidly increases with tap number to 17, after which it levels off, and so 17 - 20 is the optimum tap number.

Figure 9 (ii)-(iii) shows theory and experimental results for the FHT with orders from $0.166 - 0.833$ (phase shift 15° − 90°). The FHT normalized magnitude and phase frequency response and temporal response (Fig. (iv)) are shown. The amplitude variation is < +/- 1.5dB from 480MHz to 16.45GHz with phase variation < ±0.07 rad within the passband. Experiments for real-time FHTs with Gaussian pulses were also done. Fig. 9 (iv) shows a pulse after processing by the FHT together with theory. The resulting parameters are shown in Table I of [126]. We see that the experimental curves match the theory with good performance.

The more recent FHT results with soliton crystals and 49-GHz FSR yield much better results than the earlier 200-GHz HT in bandwidth because of the larger tap number. However, it is not always true that smaller FSRs are better. The 200-GHz devices can achieve extremely high RF bandwidths of 100GHz whereas 49-GHz devices are restricted to ~ 25GHz. In summary, our results confirm that microcombs are very effective for high-bandwidth reconfigurable HTs and FHTs with reduced footprint, complexity, and cost.

## V. DIFFERENTIATION

The fundamental operation of differentiation is a key function in RF and microwave signal processing. It has wide applications for RF spectral analysis, ultra-wideband (UWB) generation, and RF spectral filters [167-171]. Integral RF differentiators using photonic techniques [172-184] including self and cross-phase modulation (SPM, XPM) [172,173] [177-179] methods, and cross-gain modulation in semiconductor optical amplifiers (SOA) [176] have all been reported. Fractional differentiators (FDs) have unique capabilities over integral differentiators [127,182,183] and have played important roles in electricity and electronics, chemistry, biology, mechanics, and even economics. Their most well-known applications have been mechatronics, control theory, and particularly image edge detection [182,183]. Despite these many applications, photonic fractional differentiators have received comparatively little attention.

Both fractional and integral photonic differentiators have centred around differentiation of the complex optical field, as opposed to the RF signal. Although successful, a photonic differentiator using a dual-drive MZM with an RF delay line [174] was limited in speed by the RF delay line bandwidth, and while those based on optical filters [175] have featured speeds up to 40-Gb/s, they typically have a fixed integral differentiation order and lack reconfigurability and flexibility. Transversal schemes for highly reconfigurable differentiators using discrete laser arrays [169, 180, 181] are a powerful approach, but also have limitations including in the number of wavelengths. Increasing the number of lasers increases the system complexity, cost, and size, and limits the performance. A single source that can generate many wavelengths would be highly advantageous. A photonic RF integral differentiator with 8 taps [109] using a 200-GHz FSR micro-comb source recently achieved 1st, 2nd, and 3rd order derivatives at RF bandwidths up to 17 GHz, with the potential to reach 100 GHz.

A temporal fractional $N_{th}$-order differentiator is a linear and time-invariant system with transfer function [168]

$$H_N(\omega) \propto (j\omega)^N \qquad (5)$$

where $j = \sqrt{-1}$, $\omega$ is the angular RF frequency, and $N$ is the fractional or even complex [184,185] order of the derivative. According to Eq. 5, the temporal differentiator amplitude response is proportional to $|\omega|^N$, versus the phase response that is linear with a phase step of $N\pi$ at the null frequency.

Transverse filter based photonic RF differentiators have a finite set of weighted and delayed replicas of the RF signal in the optical domain that are combined by photodetection. The transfer function (Eq. (5)) is

$$H_N(\omega) = \sum_{n=0}^{M-1} a_n e^{-j\omega nT} \qquad (6)$$

where $a_n$ is the tap coefficient of the $n^{th}$ tap, $T$ is the delay time between taps and $M$ is the tap number. Differentiators described by Eq. (6) are RF intensity differentiators where the output signal after detection is a derivative of the RF signal, in stark contrast to optical field differentiators that generate the complex optical field derivative [167,171,153,154,185-187].

Recently, [109] the first RF differentiator based on a micro-comb was reported, achieving integral 1st, 2nd, and 3rd order derivatives of the RF signal. That device was based on a 200GHz spaced Kerr micro-comb with 8, 6, and 6 taps and had a dynamic range of ~30 dB. More taps would be needed to increase the differentiation order, so that increasing the order of differentiation for a fixed number of taps also increases the required power dynamic range. To obtain better performance for a fixed number of taps, the bandwidth of the integral differentiators was decreased to ½ the Nyquist frequency. To implement Eq. (5) for a temporal differentiator, tap weights (Eq. (6)) were calculated with the Remez algorithm [188] and are listed in Table I of [109] for 1st, 2nd, and 3rd order derivatives. For other RF functions, the accuracy of the comb shaping was improved via real-time feedback control. The phase and amplitude response of 1st, 2nd, and 3rd order differentiators are plot in Figs. 10 (a)–(c) versus tap number. As the tap number increases, the error between the perfect and actual theoretical amplitude response is improved for all orders. In contrast, the phase response is the same as the ideal differentiator independent of the tap number.

The shaped optical comb spectra shown in Figs. 11 (b)–(d) reveal a good match between the measured comb lines' power (red solid line) and the calculated ideal tap weights (green crosses). Figures 12 (a-i), (b-i), and (c-i) show the measured and calculated amplitude response of the differentiators. The corresponding phase response is depicted in Figs. 12 (a-ii), (b-ii), and (c-ii) where it is clear that all three differentiators agree well with theory. The periodicity of the RF response was ~16.9 GHz, given by the delay time between taps. By adjusting this periodicity by programming tap coefficients changing the fibre delay, a variable bandwidth can be achieved, which is useful for different applications. The ability to change the derivative order just by changing tap weights is a powerful method reconfigure the system functionality dynamically without new hardware, and is not available with other approaches such as optical filters.

Real-time differentiators have also been achieved for ~120 ps pulses (Fig. 13 (a)) shown in Figs. 13 (b)–(d) for different orders, along with theory, showing good agreement. Unlike optical field differentiators [167,171,153,154,185-187], RF intensity derivatives are achieved for 1st, 2nd, and 3rd order differentiators with RMSE's of ~4.15%, ~6.38%, and ~7.24%, respectively.

Tan *et.al* recently [127] achieved the first photonic fractional-order RF differentiator operating on the RF signal and not the complex optical field, using a 49-GHz-FSR Kerr micro-comb that generated a large number of wavelengths (80 in the C band), yielding a wide bandwidth of 15.49 GHz. By shaping the comb lines with the required weights, dynamically changeable fractional orders from 0.15 - 0.9 were achieved. Real-time fractional differentiation of input pulses were also performed, achieving good agreement with theory. This confirms this an effective way of implementing reconfigurable high-speed fractional differentiators with reduced complexity, footprint, and even possibly cost.

The success of the differentiator was due to the soliton crystals as well as a low 48.9 GHz FSR, that together provided > 80 wavelengths over the C-band. The system performance was helped by the soliton crystal's robust and stable operation, easy initiation and its high efficiency. The relationship between the tap number and performance was also investigated. Figure 14 shows the theoretical fractional order differentiator transfer function for 6 orders and different tap numbers, along with theory. As seen, the RF differentiator bandwidth agree well with the agreement getting better with increasing number of taps. Figure 15 (a) shows that the tap number has a large effect on the fractional differentiator bandwidth, as seen in the experiments for a fractional order of 0.45 (Figs. 15 (b) and (c)). Increasing the number of taps to 27 increased the bandwidth to 15.49 GHz, > 91% of the Nyquist band. As many as 29 wavelengths were experimentally used for the taps, which can be increased by improving the optical signal-to-noise ratio limited by the optical amplifier noise.

The comb spectra, shaped to produce the derivatives (Fig. 16), show that the optical power for each comb line agrees well with the designed tap weights. The fractional differentiator frequency response was measured with a vector network analyzer, showing that the power response had a bandwidth of DC to 15.49 GHz. The achieved range of fractional orders (Fig. 17) resulted from the close agreement between experiment and theory for the slope coefficients of the power response and shift in the phase response. Figure 18 shows the temporal response of the fractional differentiator for all orders, using a ~200ps wide RF Gaussian pulse (Fig. 18) generated by an arbitrary waveform generator. The good agreement with theory indicates the achievable range of orders that are possible with our approach to fractional differentiators.

## VI. INTEGRATION

Temporal integration is a key capability for signal processing systems and photonic based methods offer many advantages including wider bandwidths than electrical integrators along with a strong EMI immunity and low loss [79,80,189]. Many photonic integrators have been reported including grating [190-192] and micro-ring resonator [15,42,193] based approaches. These have achieved time resolutions as short as 8 ps [193] with large time-bandwidth products that result from very high Q factor resonators. These approaches, however, are still limited in many ways, including the fact that their parameters can't be varied dynamically including the time resolution or integration window, limiting their ability to process different RF signals. More significantly, these operate on the optical field, not RF signals themselves unless they have electrical to optical conversion which limits the bandwidth. Another important approach uses transversal filters that offer accuracy and reconfigurability resulting from parallelism, where each path is independently controlled [194-196]. By tailoring the delay, RF integration with a variable bandwidth can be realized [194-196]. Still, these have limitations due to the limited channel number for electro-optic combs or diode laser arrays, for example. These approaches have trade-offs between wavelength number and system complexity and ultimately limit the number of channels, in turn restricting the time-bandwidth product.

Kerr optical micro-combs in integrated platforms [38, 47, 68-70,118,197] offer advantages for RF integration, including high wavelength numbers and reduced footprint and complexity. Recently, a reconfigurable temporal RF photonic integrator was reported using a soliton crystal micro-comb [123] where the RF signal was multicast onto the comb lines, progressively delayed and then combined via photodetection. The large number of wavelengths (81) in that work yielded a large time integration window of ~6.8 ns and resolution of ~84 ps.

Figure 19 shows the transversal filter photonic RF integrator operation principle, where integration is achieved by a discrete convolution time-spectrum operation between the RF signal $f(t)$ and the micro-comb, described by:

$$y(t) = \sum_{k=1}^{N} f(t+k \cdot \Delta t)$$

(7)

where $N$ is the number of wavelengths and the delay step $\Delta t$, $j = \sqrt{-1}$. After the $f(t)$ replicas are progressively delayed and combined, the integration of $f(t)$ results with a time resolution of [194] $\Delta t$, an bandwidth of $1/\Delta t$, and integration time window $T = N \times \Delta t$. The transmission frequency response of a continuous, or non-discrete, ideal integrator is linear with $1/(j\omega)$, where $\omega$ is the angular frequency, and is a low pass sinc filter [111,117], given by

$$H(\omega) = \sum_{n=0}^{N-1} e^{-j\omega n \Delta t}$$

(8)

The experiment for the RF photonic integrator is similar to Figure 4, but based on the 49 GHz soliton crystal micro-comb that featured very low intensity noise - it has even been used as a local oscillator [120]. Thus, the noise of the micro-comb did not deteriorate the integrator noise performance. The RF signal was multicast onto the micro-comb lines flattened by a WaveShaper via an electro-optical modulator (EOM). The replicas were progressively delayed via transmission through 13km of standard optical fibre and summed with a high-speed photodetector. The adjacent wavelength delay ($\Delta t$) was given by the dispersion and length of the fibre and spacing between comb lines, and this determined the integrator resolution [194]. The temporal resolution was ~ 84 ps which can be made arbitrarily small by reducing the delay line dispersion, although with a decrease in the integration window. Sixty lines in the C-band were used, resulting in a time window ($T = N \times \Delta t$) of 60 × 84 ps = 5.04 ns (Fig. 20) with a bandwidth of $1/\Delta t$ =11.9 GHz. Figs. 20(a)−(c) show the results with different Gaussian pulse RF signals with a width varying from 0.20 ns to 0.94 ns, and a time window $T$ (~5 ns) that agrees with theory (5.04 ns). Figures 20(d)−(e) show the integration of dual Gaussian pulses separated by 1.52 ns and 3.06 ns, where the integrator displays three distinct intensity steps. The left step corresponds to integration of the first pulse while the middle step is the integration of both pulses and the right step is the integration of only the second pulse, since that pulse lies outside the time window of the first pulse. The integrator was then tested with a rectangular input waveform with a width equal to the integration window (5 ns). The measured integrated waveform exhibited a triangular shape that agreed with theory.

Figure 20 shows small discrepancies between experiment and theory - a result of the non-ideal impulse response of the system arising from small effects such as the wavelength variation of the optical amplifier gain, and the photodetector and modulator responses. To verify this, the system impulse response to a Gaussian pulse input was measured. Since the system time resolution (~ 84 ps) was much less than the pulse width, the wavelength channels were separated into multiple subsets, each with a much larger spacing between the adjacent comb lines and so obtaining a temporal resolution larger than the input pulse duration, and their impulse response was then sequentially measured. Figure 21(a) shows the measured system impulse response which was not flat even if the comb lines were perfectly uniform. The RF signal and impulse response in Fig. 20 were used to calculate the integral output, with the results matching well with experiment, indicating that errors were induced by the non-ideal system impulse response. These errors are in fact easy to compensate for by, for example, using more accurate comb shaping where the feedback loop error signal is generated directly by the measured impulse response and not the optical power of the comb lines. Hence, this implicitly compensated for the wavelength dependence of all components. A flat impulse response resulted, (Fig. 21(b)), much closer to the ideal response than Fig. 21(a). Integration was then performed for the same RF signals and the results are shown in Fig. 22. During these measurements, 81 wavelength channels were used by the impulse response shaping process, yielding an time integration window ($T = N \times \Delta t$) that was increased to 81 × 84 ps = 6.804 ns, resulting in a bandwidth of 1/84ps = 11.9 GHz and time-bandwidth product of 6.804 ns × 11.9 GHz = ~81 (approximately equal to $N$, the number of channels). The measured integrals (Fig. 22) show much better agreement with theory, indicating the advantage of the shaping impulse response compensating method as well as the overall power and feasibility of our RF integrator.

## VII. Discussion

We have reviewed recent progress on the use of Kerr micro-combs for high bandwidth temporal RF and microwave signal processing, focusing on integral and fractional order Hilbert transformers, high order and fractional differentiators, both with continuously tunability in their order, and finally RF integrators. Micro-combs produce a large number of comb lines that greatly increase the performance and processing bandwidth of RF systems in both amplitude and phase response. They have achieved powerful results for RF signal processing because of their large number of comb lines, compact footprint, and large comb spacings compared to other approaches such as electro-optic combs. The devices achieved operation bandwidths of ~ 16 GHz, for the 49 GHz devices and ~ 25 GHz for 200GHz FSR combs, evaluated in the frequency domain with Vector Network Analyzers and time domain with Gaussian pulses.

The performance of frequency domain signal processors including Hilbert transformers and differentiators is determined by the number of wavelengths, or taps. We have reviewed the benefit of Kerr micro-combs for photonic RF and microwave transversal filters, showing the contrast between using widely spaced versus low spaced micro-combs – ie., 200GHz versus 49 GHz combs, in performance. The 80 lines supplied by the 49GHz comb (vs 20 for the 200-GHz comb) produced much better performance. In contrast, the 49-GHz device has a lower bandwidth, being limited to the 25 GHz Nyquist zone, while the 200-GHz devices had RF bandwidths well beyond standard electronic microwave technology.

Two types of combs employed to realize RF functions were qualitatively very different. Apart from the spacings being very different, the widely spaced 200GHz FSR microcomb operated in a quasi-coherent state that did not feature solitons, whereas the low spaced microcomb with an FSR of 49GHz operated in a state referred to as soliton crystals. Within the optical C band, the 49GHz spaced microcomb enabled up to 90 comb lines, yielding significantly better performance for the temporal signal processors than the than the 200GHz spaced microcomb which only allowed 20 taps in the C-band. Further, the soliton crystal states offered many other advantages such as much lower noise, higher stability, and much easier generation including the ability to even initiate soliton crystals through simple manual tuning of the pump wavelength. However, on the other hand the 49GHz combs generally yielded much smaller Nyquist bandwidths typically near ~16 GHz. For the transversal temporal signal processors such as the integrators, the large number of wavelengths brought about a large number of broadcasted RF replicas, thus yielding a large integration window for the integrator. The temporal response of the integrator was measured with a diverse range of RF inputs, verifying a integration window of 6.8 ns and a time feature as fast as 84 ps. We achieve excellent agreement of our experimental results with theory, confirming that our approach is effective for reconfigurable high-speed signal processing with high bandwidths for microwave systems.

Micro-combs offer many benefits for photonic RF signal processing. First, the coherent soliton states will enable advanced RF functions such as clock generation and wideband frequency conversion. Maturing nanofabrication techniques can achieve high Q MRRs for comb generation with FSRs from 10s' of GHz [119] to THz [47], covering the full RF bands of interest for almost any integrated RF system. The wide range in FSRs of micro-combs yields a large range in Nyquist zone to 100's of GHz, well beyond electronics and also challenging even for mode-locked lasers and EO modulation OFCs. Using tailored dispersion, micro-combs with ultra-wideband bandwidths - even octave-spanning, [71] can be achieved, enabling a very large number of wavelengths using only a single integrated source, to work in conjunction with broadband opto-electronic equipment. This approach enables significantly enhanced wavelength-division parallelism for massive data transmission and processing [62-64] for many applications including radio-over-fibre systems, for example.

Perhaps most importantly, the high reconfigurability of micro-comb based programmable RF photonic transversal filters enables one to achieve highly versatile processing functions with dynamic real-time reconfigurability so that the same system can be dynamically reprogrammed to perform a wide range of different functions without having to alter any hardware. Simply by programming the comb line intensities according to the required tap weights, the same hardware can achieve many different signal processing functions. This very high level of reconfigurability is unprecedented for photonic RF signal processors and enables highly reconfigurable processing functions and high processing accuracy that typically cannot be obtained by passive photonic integrated circuits or via optical analogue signal processing. Further, very wide operational RF bandwidths can be achieved for diverse computing requirements of many practical applications. Photonic transversal filters are similar to filters achieved using digital electronic signal processing except realized with photonics.

There are a number of factors that can lead to tap errors during the comb shaping, thus leading to non-ideal impulse responses of the system as well as deviations between the experimental results and theory. These factors include optical micro-comb instability, waveshaper accuracy, wavelength variation in optical amplifier gain, modulator induced chirp, and dispersive fibre third-order dispersion. These errors can be reduce by using feedback control paths in real-time, where the comb line power is detected by an OSA and compared with the theoretical tap weights, generating an error signal fed back to the waveshaper for accurate comb shaping. The best way of generating the error signal is directly from the measured impulse response instead of the raw optical power of the comb lines. In this approach, the impulse response of the system is obtained by measuring replicas of an RF Gaussian pulse at all wavelengths, further extracting the peak intensities to obtain accurate RF-to-RF wavelength channel weights. After this, the measured channel weights are subtracted from the theoretical weights to obtain an error signal then used to set the Waveshaper loss. After several iterations of comb shaping an accurate impulse response that compensates the non-ideal impulse response of the system can be obtained, thus significantly improving the RF photonic transversal filter accuracy. This calibration process only needs to be done once.

Finally, although the results reviewed here relied on some benchtop components, such as the commercially available Waveshaper, RF photonic signal processors based on Kerr micro-combs have significant potential to achieve much higher levels of integration with current nanofabrication techniques. To begin with, the most important component is the micro-comb source itself, which not only is already integrated but is fabricated with CMOS compatible processes. CMOS fabrication foundries can already perform advanced hybrid integration of microcombs with III-V devices, and this will ultimately enable the monolithic integration of entire RF systems. Other key components have been realized in integrated form with state-of-art nanofabrication techniques [198-200], including the optical pump source [131, 139], LiNO$_3$ modulators [200], optical spectral shapers [198], large dispersion media [199] and photodetectors. Further, advanced integrated microcombs have been demonstrated [139] that can generate soliton crystals reliably in turn-key operation. Monolithically integrating the whole RF processing system would greatly strengthen the performance, compactness and energy efficiency of the system. Even without this, however, using the discrete integrated comb sources to replace discrete laser arrays already yields significant benefits for RF and microwave systems in terms of performance, size, cost, and complexity.

## VIII. CONCLUSION

We review progress in applications of Kerr micro-combs to RF and microwave photonic time based signal processing using transversal filters. Optical micro-combs are a new generation of multi-wavelength compact sources for the RF photonics community to exploit, offering huge benefits for high-performance RF signal processing with reduced footprint, complexity, and potentially cost. By programming the comb lines according to calculated tap weights, a wide range of signal processing functions have been experimentally demonstrated. We have focused on fractional and integral order Hilbert transforms and differentiators as well as RF integrators that operate on the RF signal rather than the complex optical field. Real-time system demonstrations are also performed that show good agreement with theory. These results verify that Kerr micro-combs are highly effective in achieving high-speed reconfigurable signal processing for future ultra-high performance RF systems

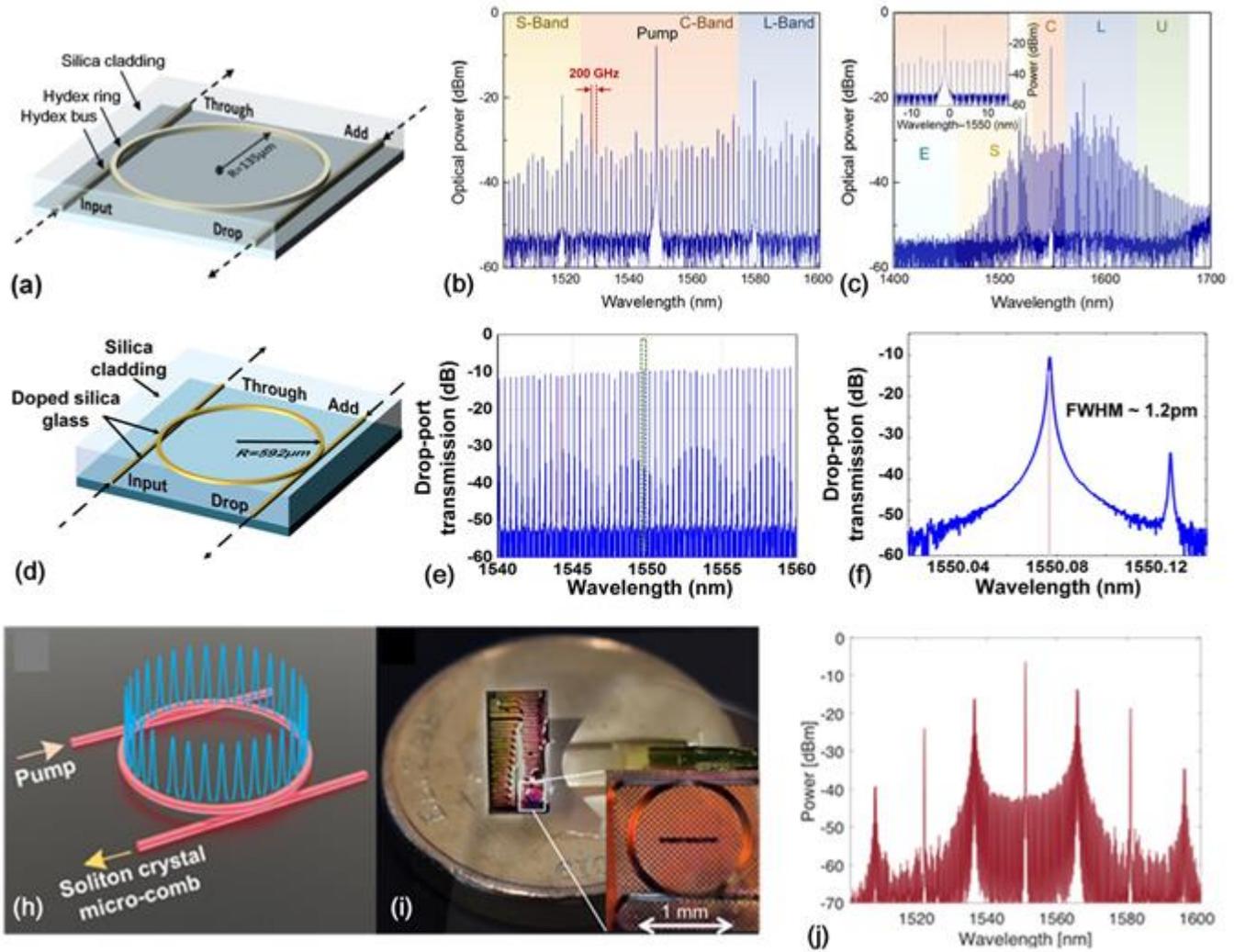

Fig. 1. Schematic illustration of the integrated MRRs for generating the Kerr micro-comb for both (a-c) 200GHz FSR combs and (d-j) 49GHz combs. (b, c) Optical spectra of the micro-combs generated by 200GHz MRR with a span of (b) 100 nm and (c) 300 nm. (j) Optical spectra of the micro-combs generated by 50GHz MRR with a span of 100 nm. (e) Drop-port transmission spectrum of the integrated MRR with a span of 5 nm, showing an optical free spectral range of 49 GHz. (f) A resonance at 193.294 THz with full width at half maximum (FWHM) of 124.94 MHz, corresponding to a quality factor of $1.549 \times 10^6$.

Figure 1

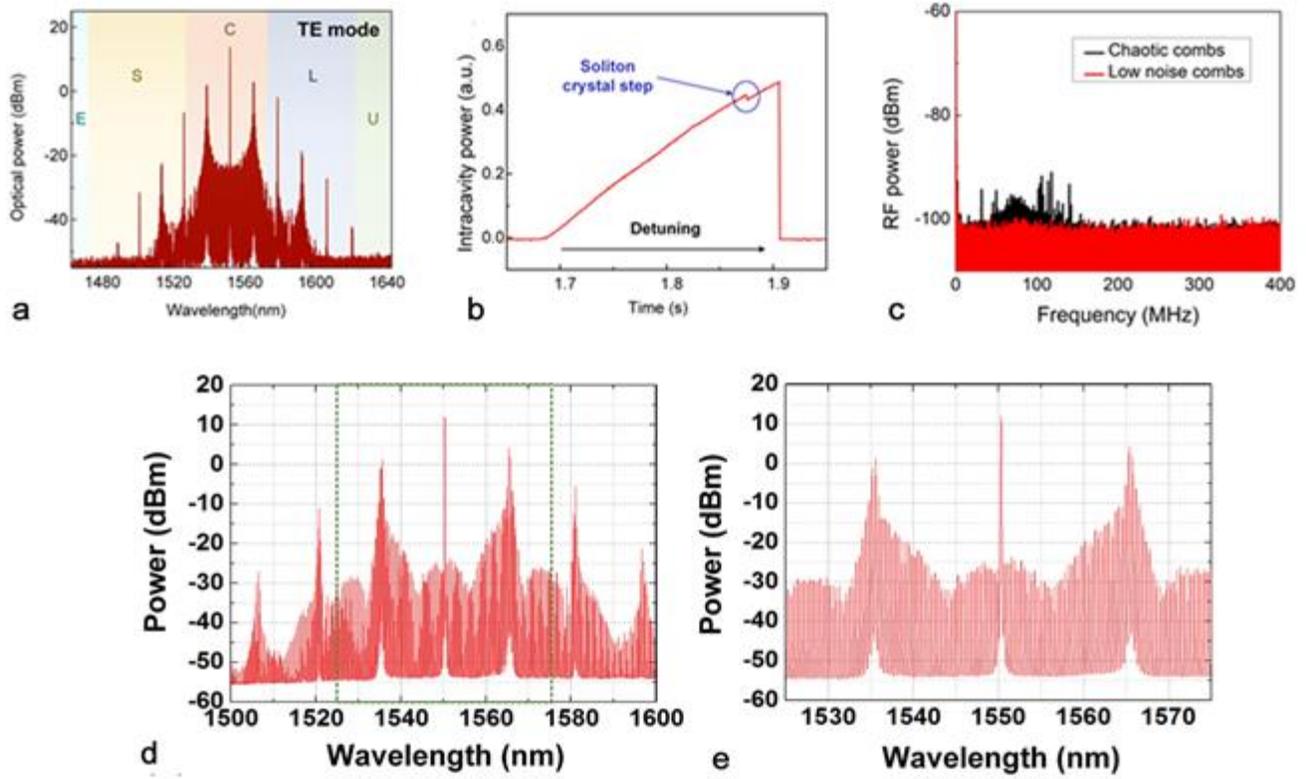

Fig. 2. (a, d, e) Optical spectra of various soliton crystal micro-combs. (b) Optical power output versus pump tuning, showing the very small power jump at the onset of soliton crystal combs. (c) Transition from high RF noise chaotic state to low noise state of the soliton crystal comb.

# Figure 2

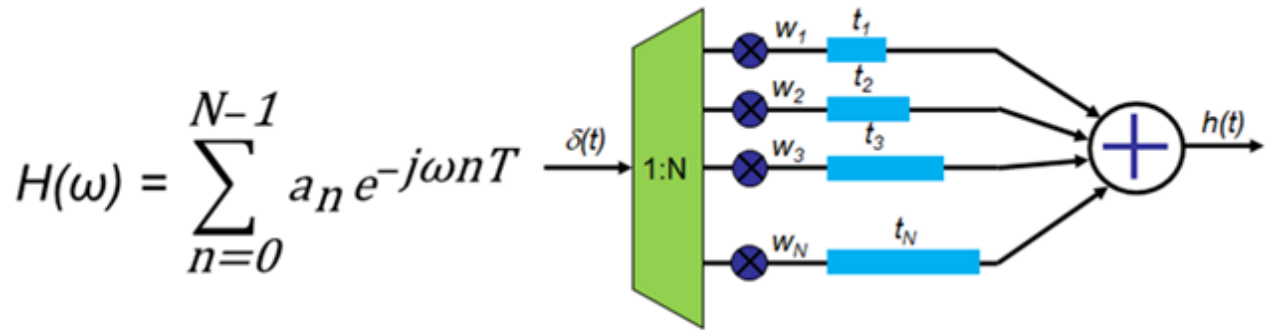

$$H(\omega) = \sum_{n=0}^{N-1} a_n\, e^{-j\omega nT}$$

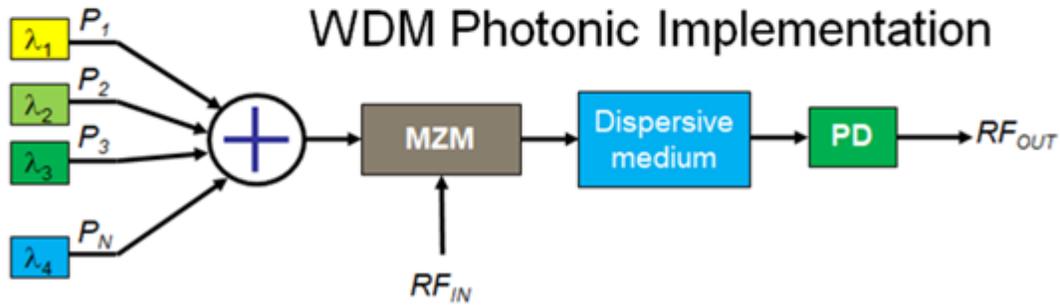

Fig. 3. Theoretical schematic of the principle of transversal filters using wavelength multiplexing. MZM: Mach-Zehnder modulators. PD: photo-detector.

Figure 3

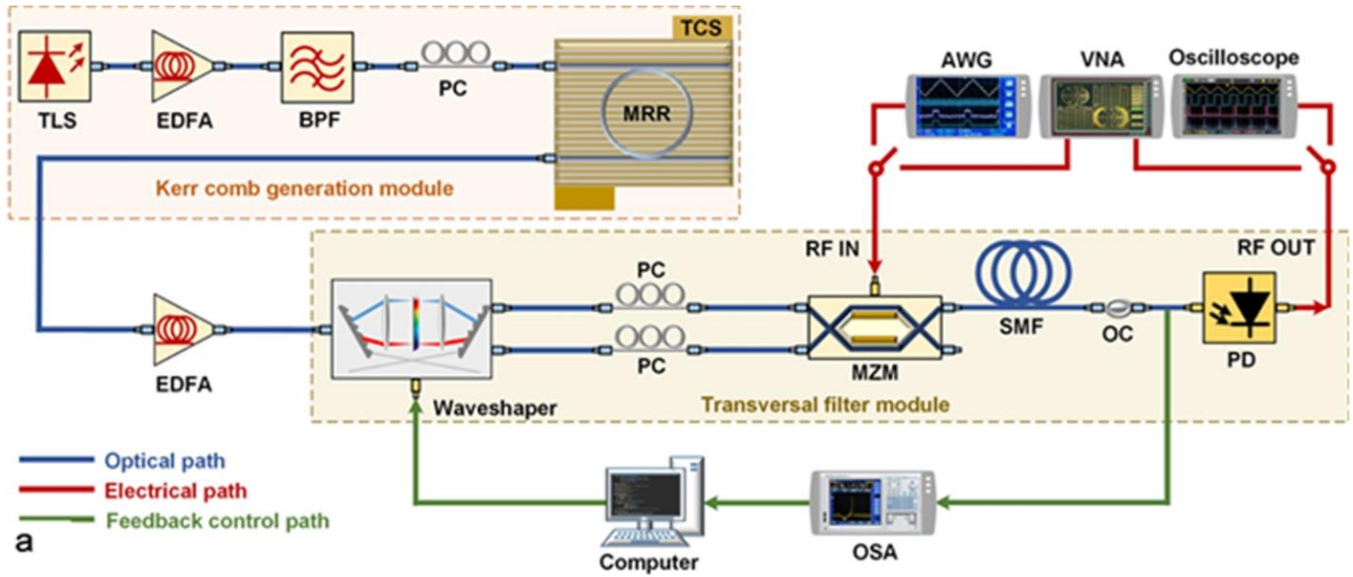

Fig. 4. Experimental schematic for RF transversal filters based on 200GHz microcomb (top) and 49GHz microcomb (bottom). TLS: tunable laser source. EDFA: erbium-doped fiber amplifier. PC: polarization controller. BPF: optical bandpass filter. TCS: temperature control stage. MRR: micro-ring resonator. WS: WaveShaper. OC: optical coupler. SMF: single mode fibre. OSA: optical spectrum analyzer. AWG: arbitrary waveform generator. VNA: vector network analyser. PD: photodetector. BPD: balanced photodetector.

Figure 4

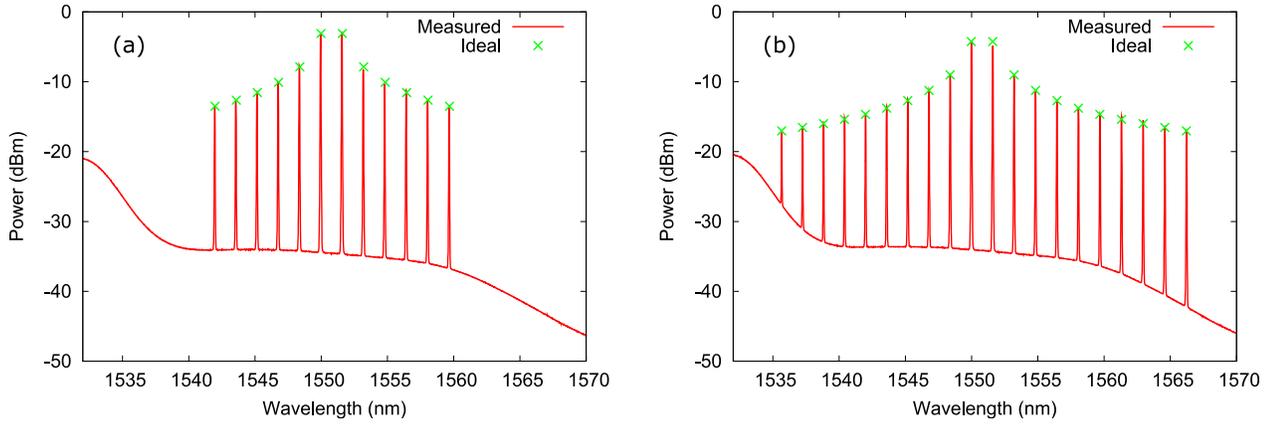

Fig. 5. Shaped optical spectra showing the weight of each tap for: (a) 12 tap filter, and (b) 20 tap filter

Figure 5

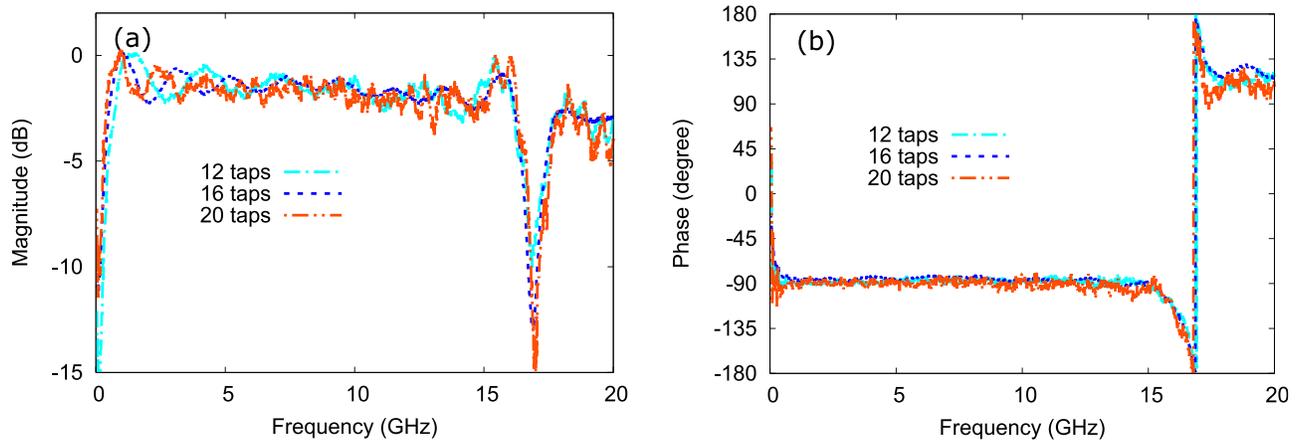

Fig. 6. Measured system RF frequency response for different number of filter taps: (a) amplitude; and (b) phase response.

Figure 6

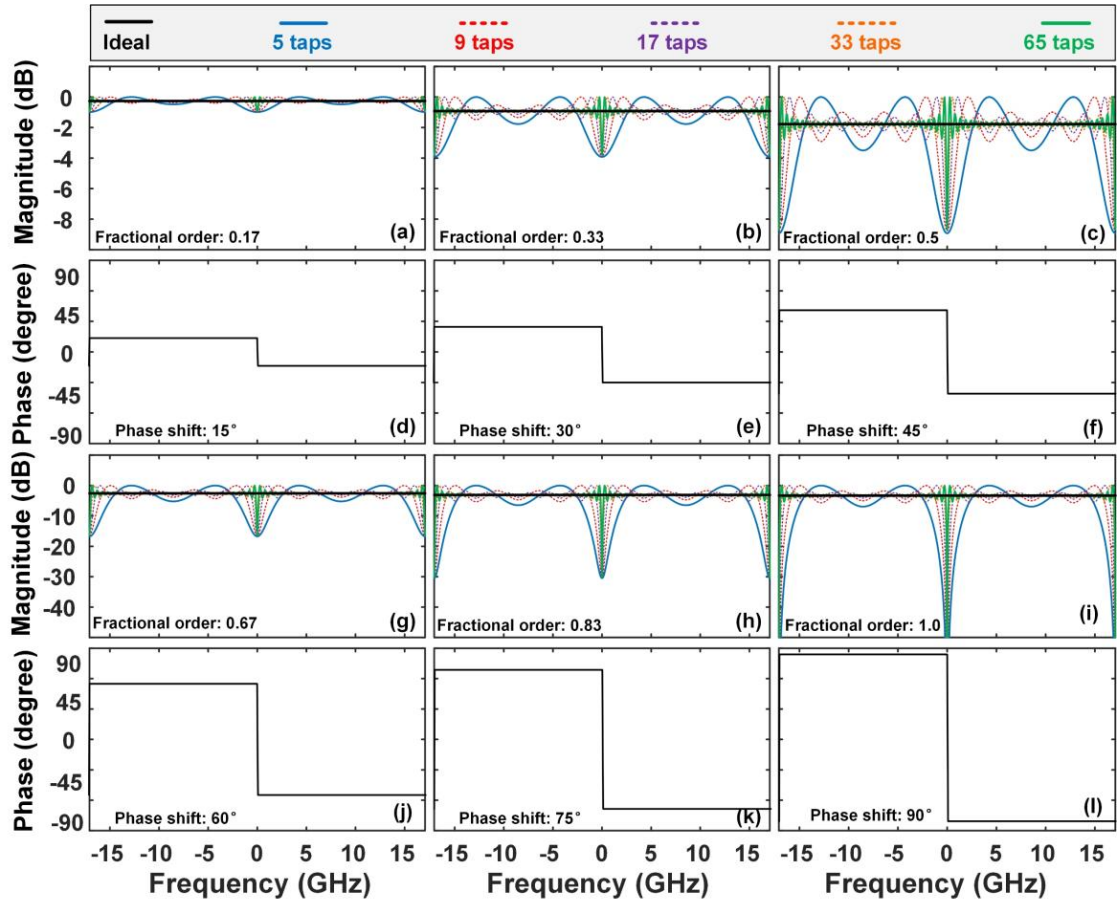

Fig. 7. Theoretical RF amplitude and phase response of FHTs with (a, d) 15º, (b, e) 30º, (c, f) 45º, (g, j) 60º, (h, k) 75º, and (i, l) 90º phase shifts. The amplitude of the fractional Hilbert transformers designed based on Eq. (3) (colour curves) are shown according to the number of taps employed.

(Maybe separate the top and bottom parts a bit)

Figure 7

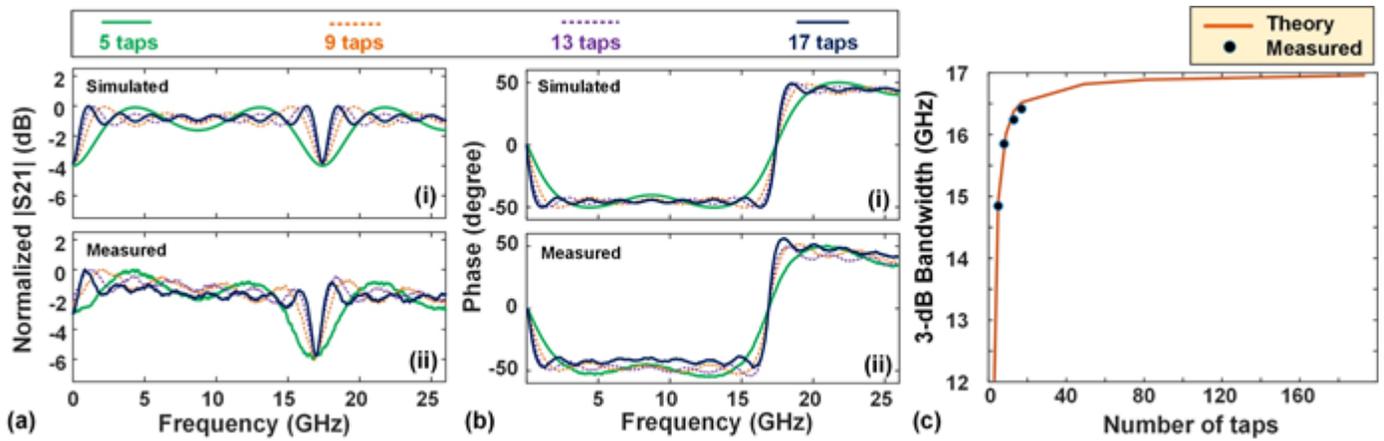

Fig. 8. (a) and (b) Simulated and measured amplitude and phase response for the FHT for different numbers of taps for a FHT phase shift of 45°. (c) Simulated and experimental results of 3-dB bandwidth with different taps for a phase shift of 45°.

Separate figure c into its own figure

Figure 8

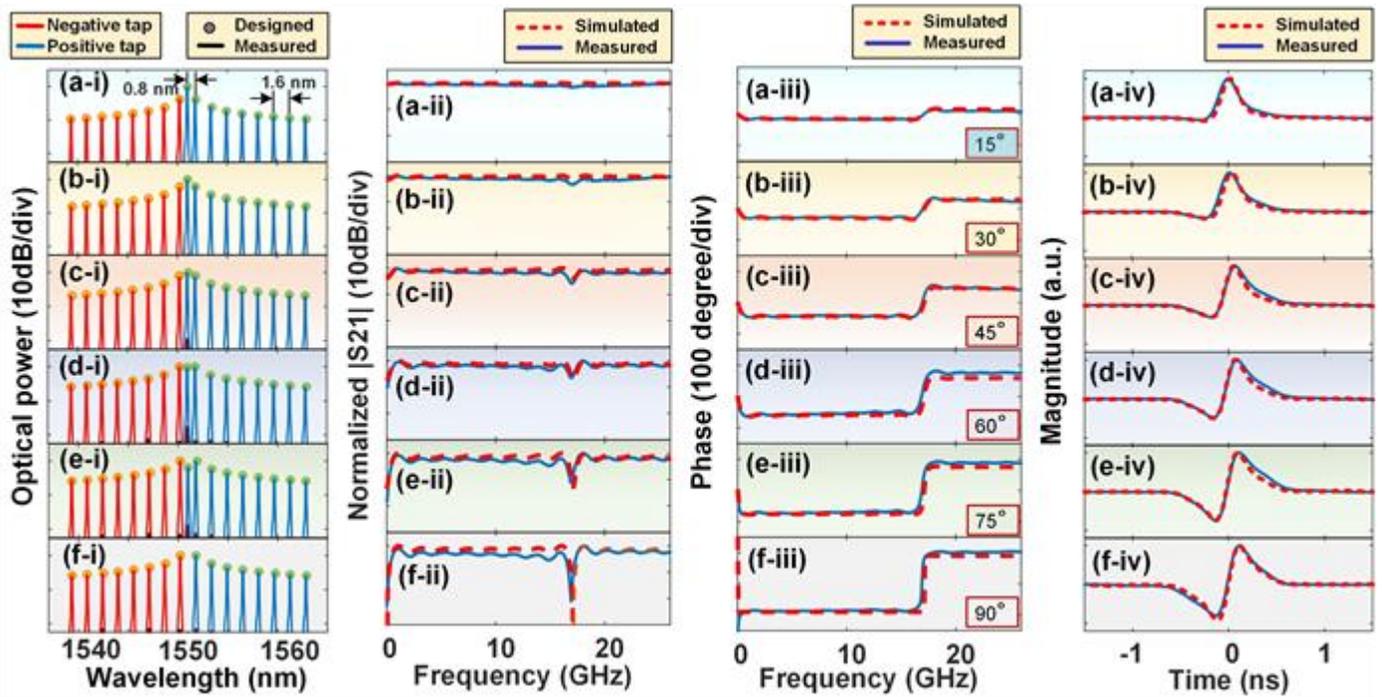

Fig. 9. Simulated (dashed curves) and experimental (solid curves) results of FHT with various phase shifts of (a) 15°, (b) 30°, (c) 45°, (d) 60°, (e) 75°, and (f) 90°. (i) Optical spectra of the shaped micro-comb corresponding with positive and negative tap weights (ii) RF amplitude responses with fractional orders of 0.166, 0.333, 0.5, 0.667, 0.833, and 1. (iii) RF phase responses with phase shifts of 15°, 30°, 45°, 60°, 75° and 90°. (iv) Output temporal intensity waveforms after the FHT.

Figure 9

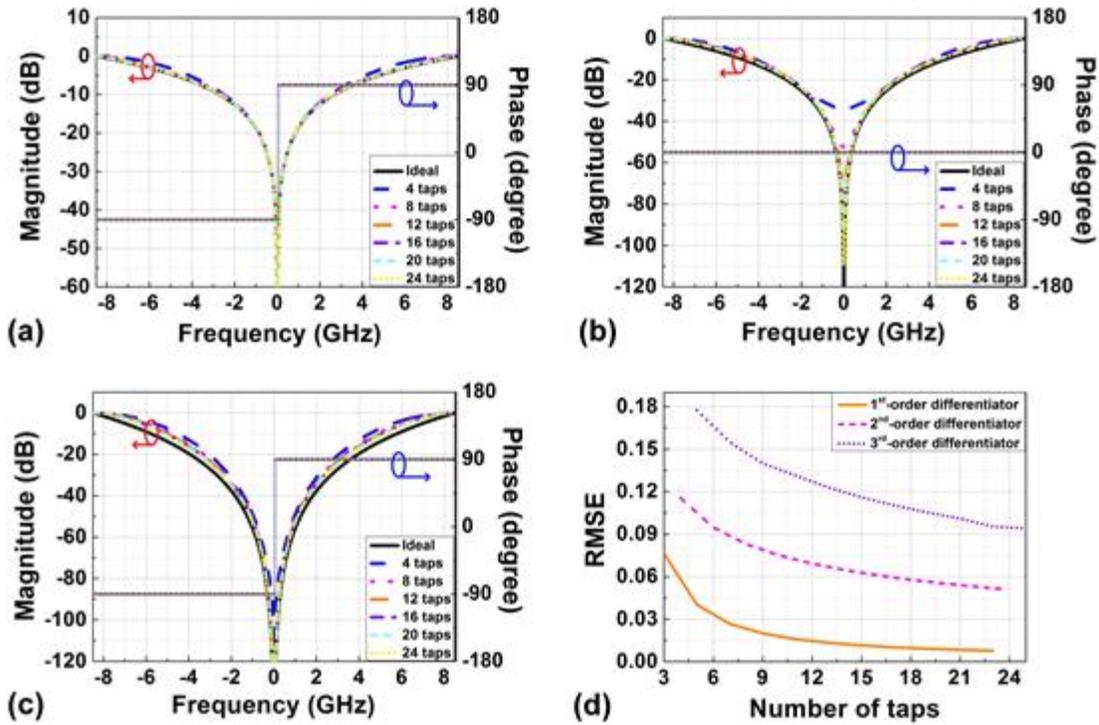

Fig. 10. Simulated RF amplitude and phase responses of the (a) first-, (b) second-, and (c) third-order temporal differentiators. (d) RMSEs between calculated and ideal RF amplitude responses of the first-, second-, and third-order intensity differentiators as a function of the number of taps.

Figure 10

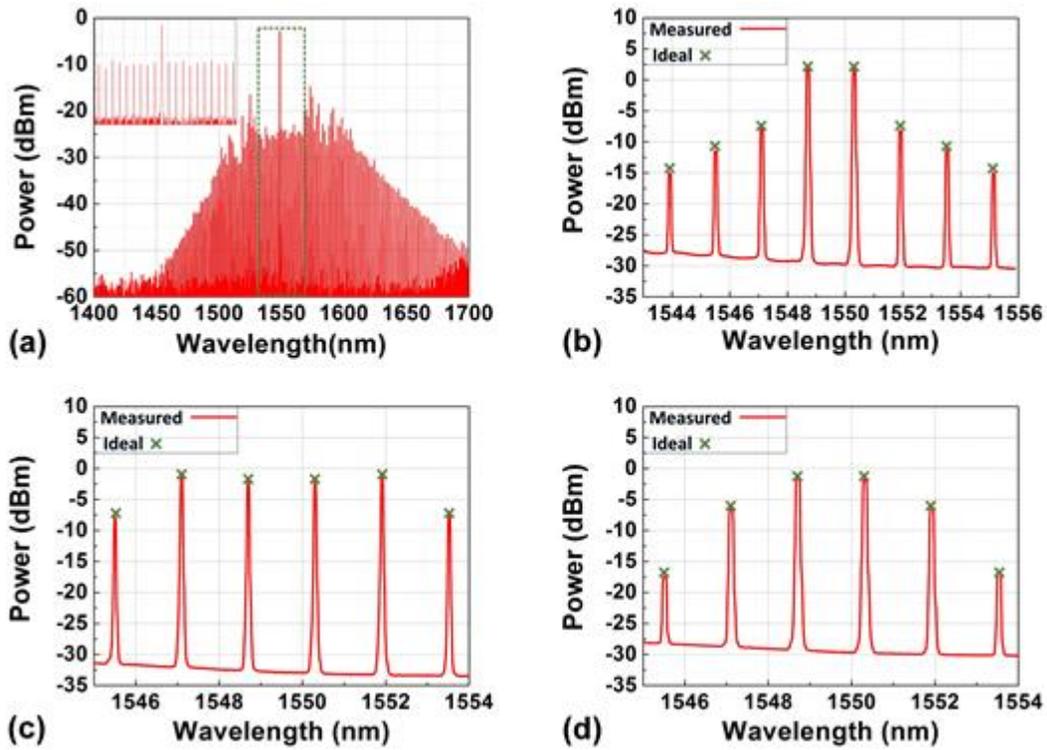

Fig. 11. (a) Optical spectrum of the generated Kerr comb in a 300-nm wavelength range. Inset shows a zoom-in spectrum with a span of ~32 nm. (b)–(d) Measured optical spectra (red solid) of the shaped optical combs and ideal tap weights (green crossing) for the first-, second-, and third-order intensity differentiators.

Figure 11

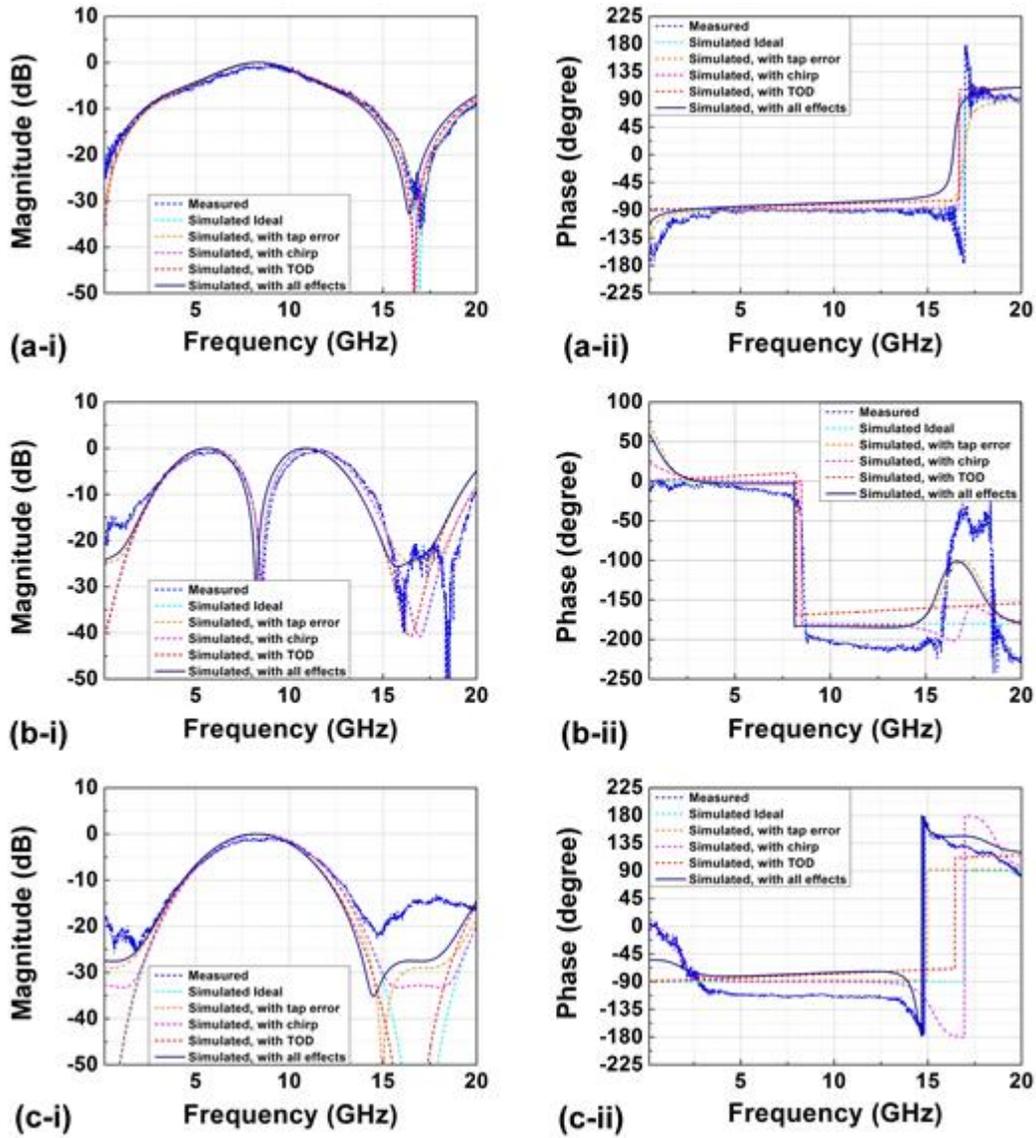

Fig. 12. Measured and simulated RF amplitude and phase responses of (a-i)–(a-ii) the first-order, (b-i)–(b-ii) second-order, and (c-i)–(c-ii) third-order intensity differentiators.

Figure 12

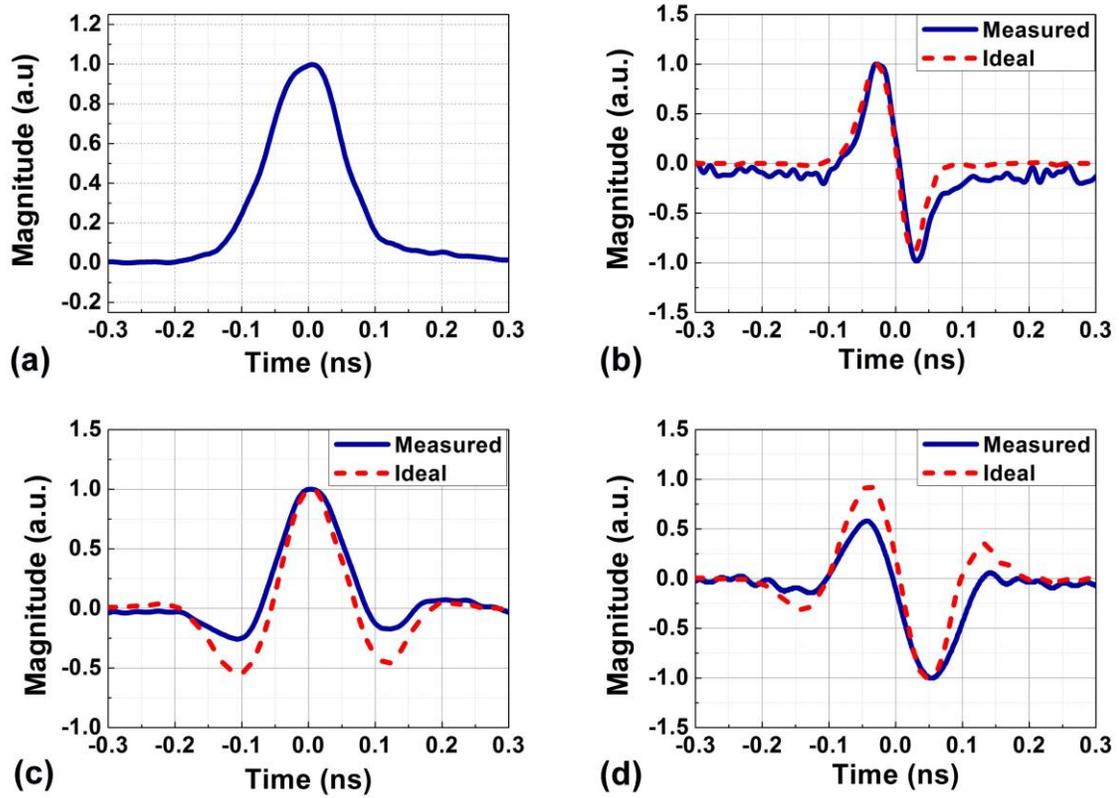

Fig.13. (a) Measured temporal waveforms of a Gaussian input pulse. Theoretical (red dashed) and experimental (blue solid) responses of the (b) first-, (b) second-, and (c) third-order differentiators.

Figure 13

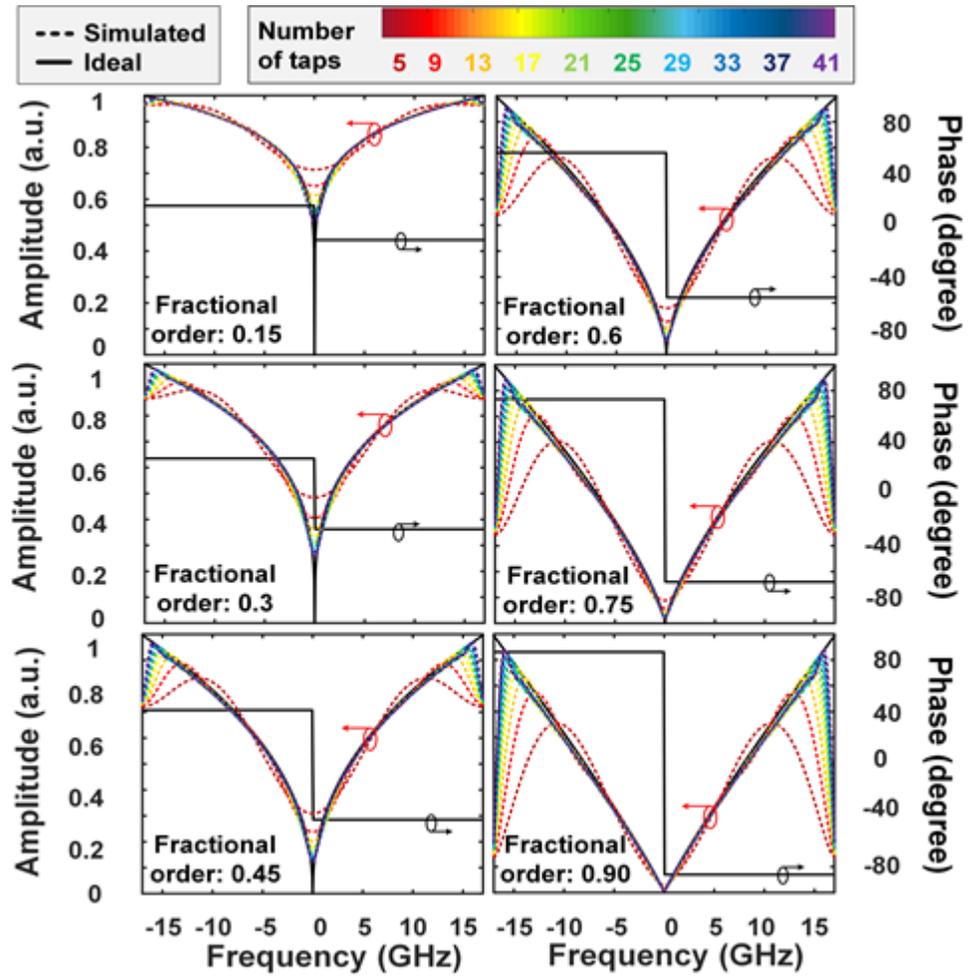

Fig. 14. Simulated transfer function of different fractional differentiation orders with varying number of taps.

Figure 14

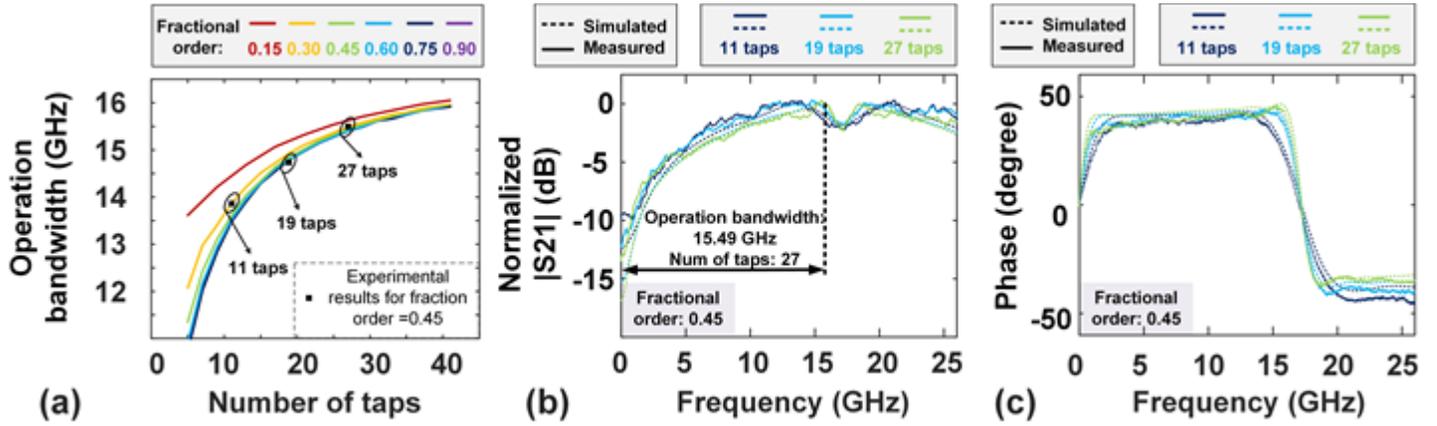

Fig. 15. (a) Relationship between the number of taps and operation bandwidth. (b, c) Experimentally demonstrated fractional differentiator with varying number of taps.

Figure 15

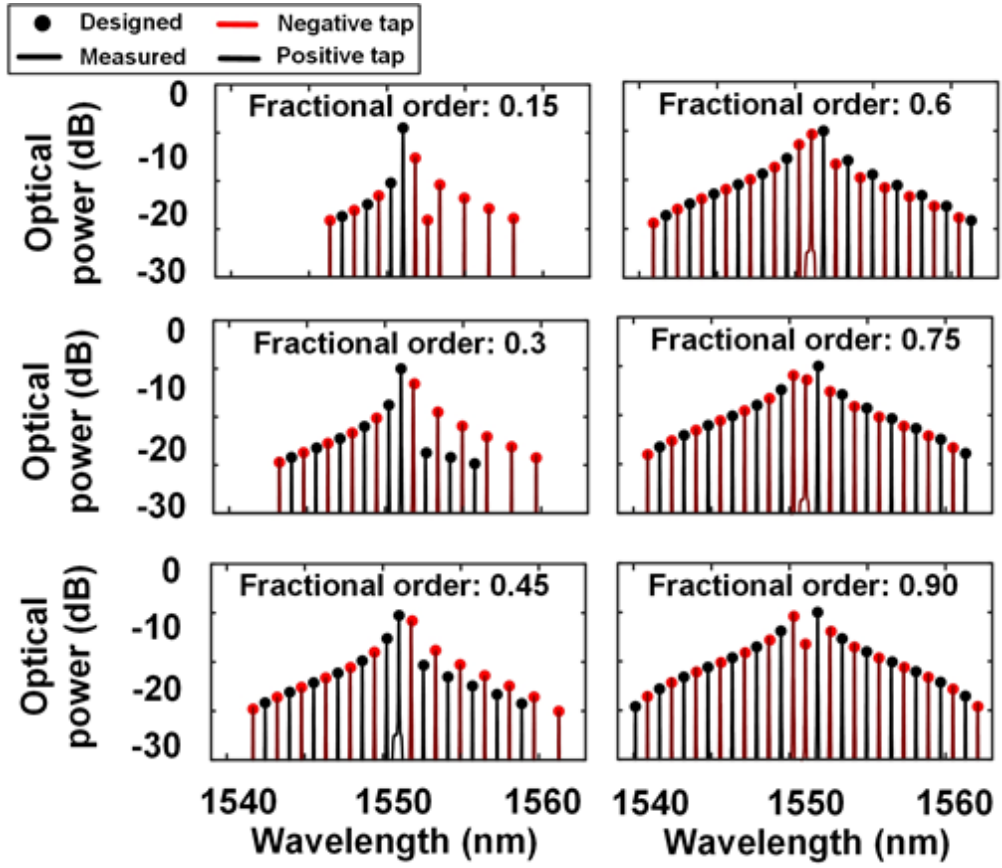

Fig. 16. Optical spectra of the shaped micro-comb for different fractional orders.

Figure 16

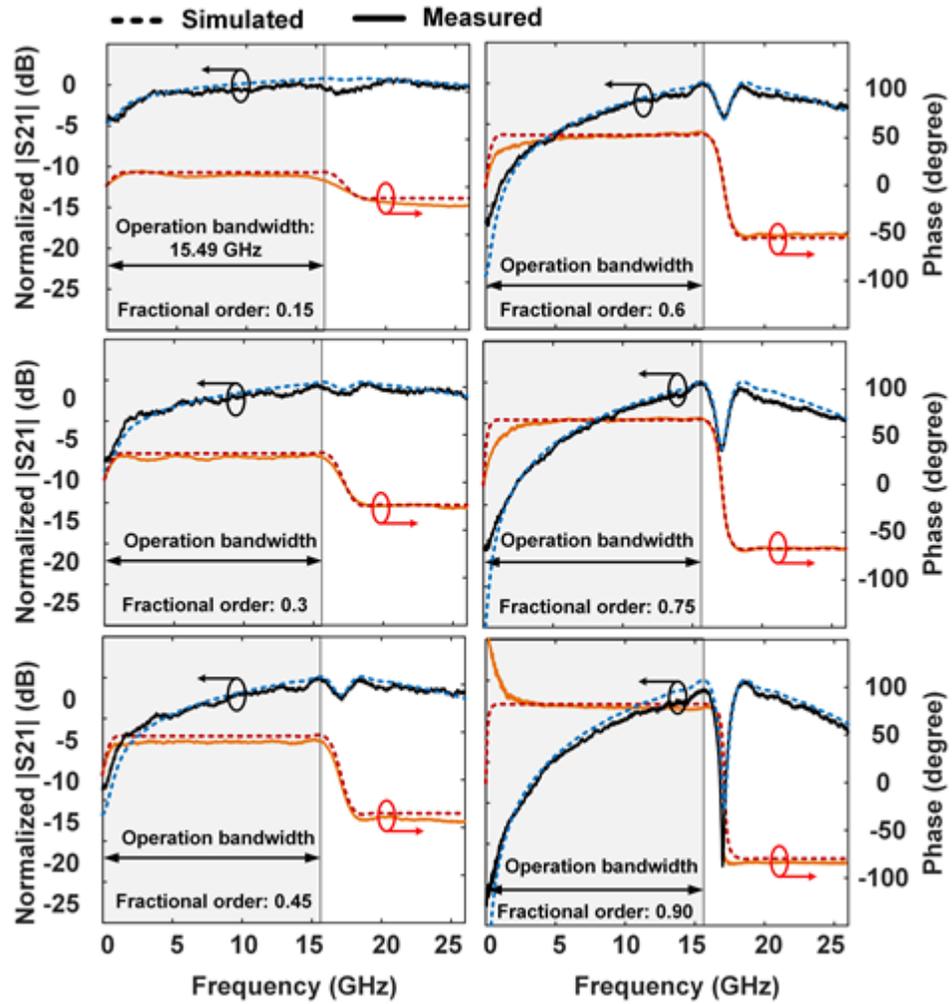

Fig. 17. Simulated and measured the transmission response of the fractional differentiator at different orders ranging from 0.15 to 0.90.

Figure 17

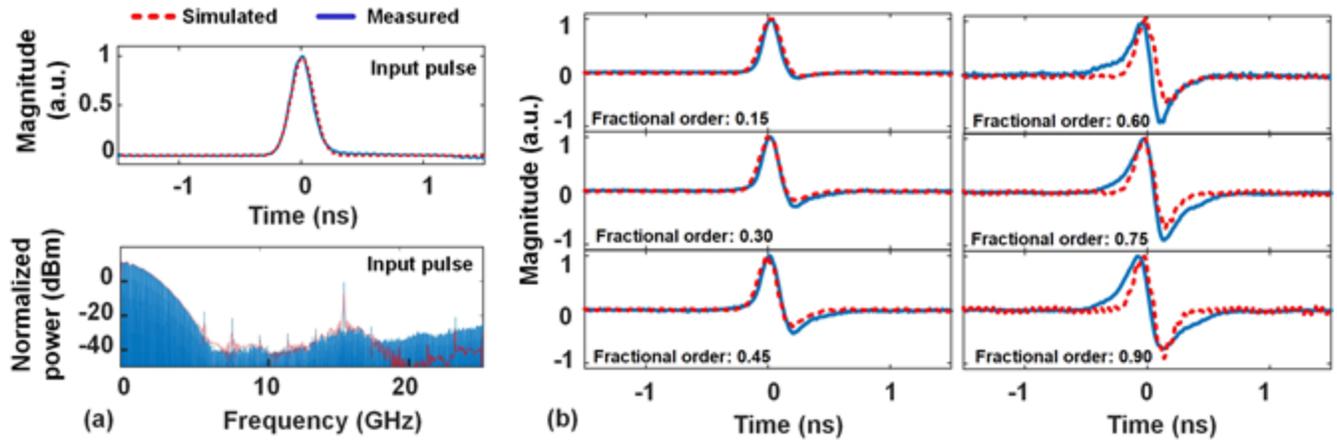

Fig. 18. Simulated and measured RF Gaussian pulse output temporal intensity waveform after the fractional differentiator.

Figure 18

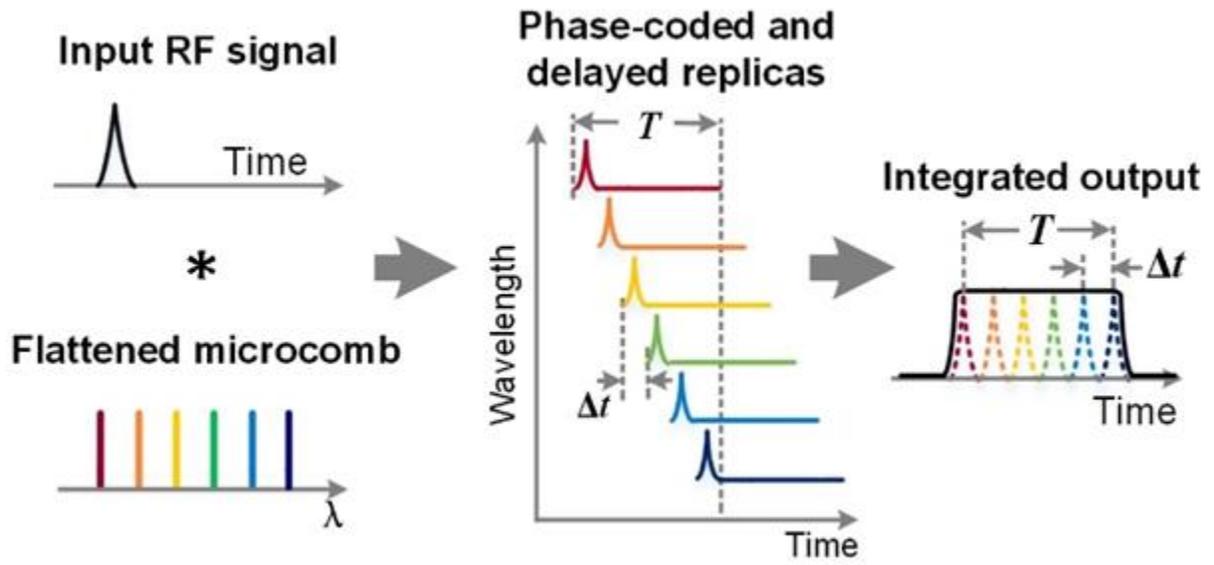

Fig. 19. Schematic diagram of the photonic RF integration.

Figure 19

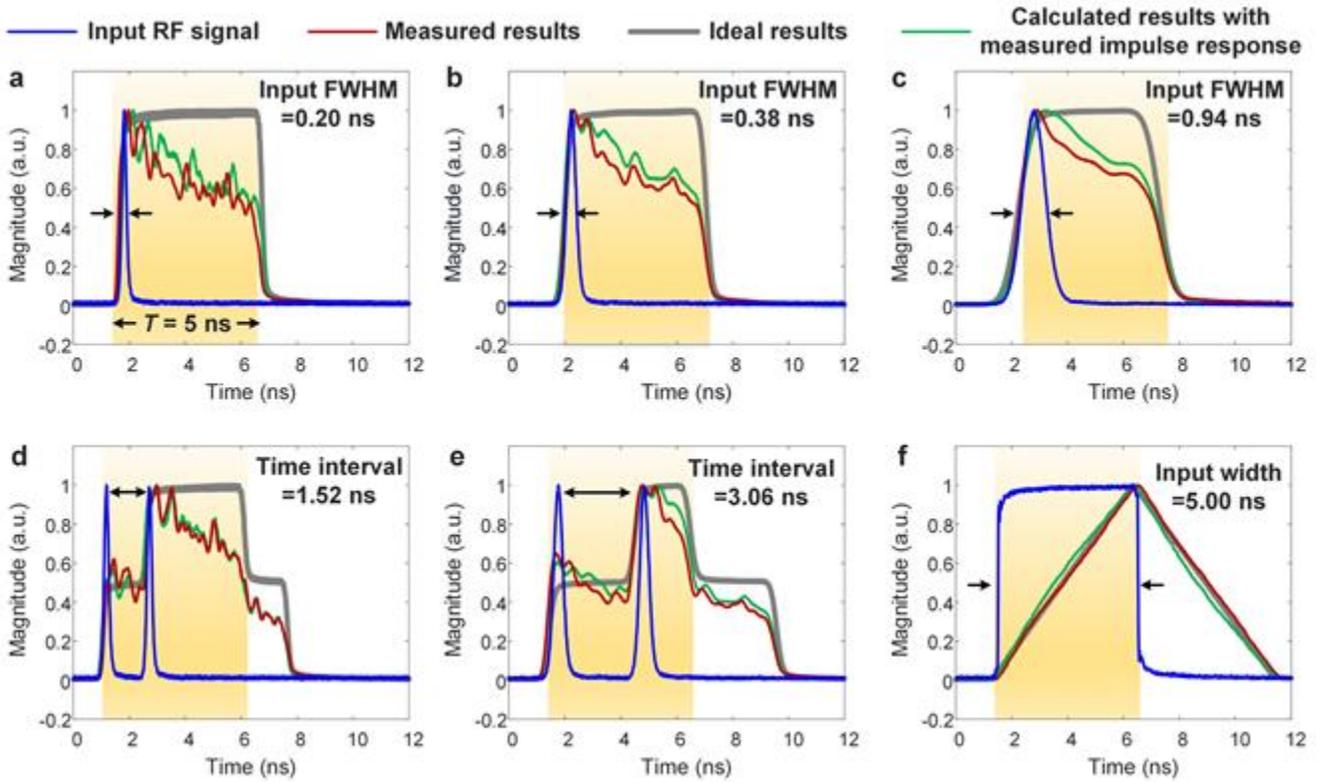

Fig. 20. Experimental results of the micro-comb-based RF integrator after comb optical power shaping for input (a-c) Gaussian pulses with FWHM of 0.20, 0.38 and 0.94 ns, (d-e) dual Gaussian pulses with time intervals of 1.52 and 30.6 ns, and (f) a triangular waveform with a width of 5.00 ns. The blue curves denote the input signal, the red curves denote the measured integration results, the gray curves denote the ideal integration results, and the green curves denote the integration results calculated with the measured impulse response of the system.

Figure 20

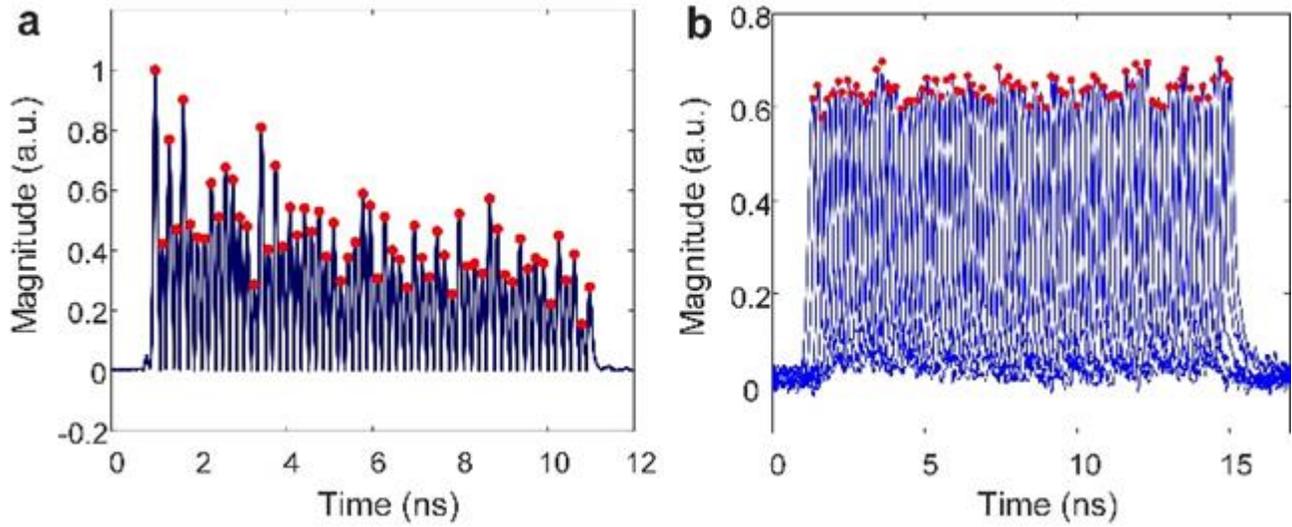

Fig. 21. Measured impulse response of the integrator (a) after comb optical power shaping and (b) after impulse response shaping using a Gaussian RF input pulse.

Figure 21

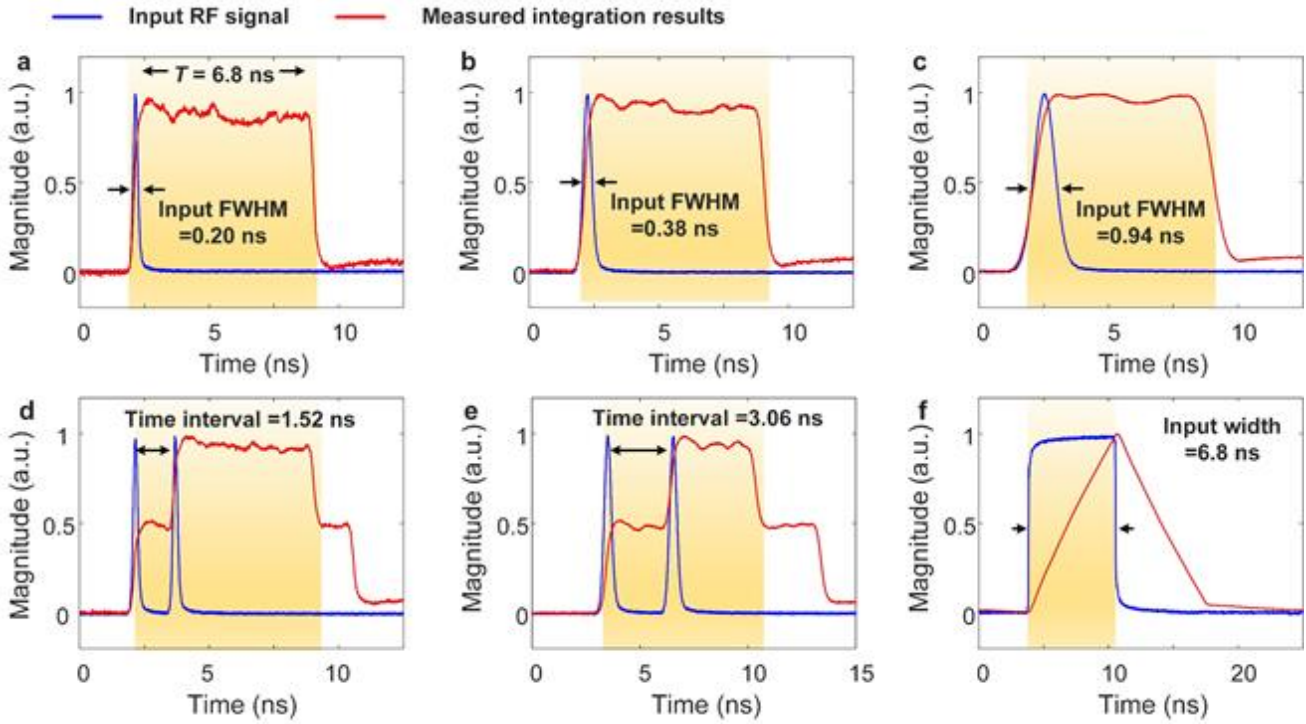

Fig. 22. Experimental results of the micro-comb-based RF integrator after impulse response shaping for input (a-c) Gaussian pulses with FWHM of 0.20, 0.38 and 0.94 ns, (d-e) dual Gaussian pulses with time intervals of 1.52 and 3.06 ns, and (f) a triangular waveform with a width of 5.00 ns. The blue curves denote the input signal, the red curves denote the measured integration results.

Figure 22